\documentclass[usenatbib]{mn2e}
\input{epsf}
\usepackage{amssymb}

\title[Mass profiles and galaxy orbits]
{Mass profiles and galaxy orbits in nearby galaxy clusters from the analysis of the projected phase-space}
\author[R. Wojtak \& E. L. {\L}okas]{Rados{\l}aw Wojtak and Ewa L. {\L}okas
\\   \\
Nicolaus Copernicus Astronomical Center, Bartycka 18, 00-716 Warsaw, Poland\\
}

\begin{document}

\maketitle

\begin{abstract}
We analyze kinematic data of $41$ nearby ($z<0.1$) relaxed galaxy clusters in terms of the projected 
phase-space density using a phenomenological, fully anisotropic model of the distribution function. We apply 
the Markov Chain Monte Carlo approach to place constraints on total mass distribution approximated 
by the universal NFW profile and the profile of the anisotropy of galaxy orbits. We find the normalization of the mean 
mass-concentration relation is $c=6.9_{-0.7}^{+0.6}$ at the virial mass $M_{\rm v}=5\times 10^{14}M_{\odot}$. By comparison 
with the calibration from cosmological $N$-body simulations it is demonstrated that this result is fully consistent 
with $\sigma_{8}$ from WMAP1 data release and agrees at $\sim1\sigma$ level with that from WMAP5. 
Assuming a one-to-one correspondence between $\sigma_{8}$ and the normalization of the mass-concentration relation in the 
framework of the concordance model we estimate the normalization of the linear power spectrum to be 
$\sigma_{8}=0.91_{-0.08}^{+0.07}$. Our constraints on the parameters of the mass profile are compared with estimates 
from X-ray observations and other methods based on galaxy kinematics. We also study correlations between the virial mass 
and different mass proxies including the velocity dispersion, the X-ray temperature and the X-ray luminosity. We demonstrate 
that the mass scaling relations with the velocity dispersion and the X-ray temperature are fully consistent with the predictions 
of the virial theorem.

%We show that the accuracy 
%of parameter estimation in our method is noticeably better than that achieved in the Jeans analysis with velocity moments 
%and comparable to that resulting from the X-ray analysis. 
%We used our mass estimates to study correlations between the virial mass and different mass proxies including the velocity dispersion, the X-ray temperature and the X-ray %luminosity. 
%It is demonstrated that the mass scaling relations with the velocity dispersion and the X-ray temperature are fully 
%consistent with the virial theorem.

We show that galaxy orbits are isotropic at the cluster centres (with the mean ratio of the radial-to-tangential 
velocity dispersions $\sigma_{r}/\sigma_{\theta}=0.97\pm0.04$) and radially anisotropic at the virial sphere 
(with the mean ratio $\sigma_{r}/\sigma_{\theta}=1.75^{+0.23}_{-0.19}$). Although the value of the central 
anisotropy appears to be universal, the anisotropy at the virial radius differs between clusters within the 
range $1\lesssim (\sigma_{r}/\sigma_{\theta})\lesssim 2$.

Utilizing the Bautz-Morgan morphological classification and information on the prominence of a cool core 
we select two subsamples of galaxy clusters corresponding to less and more advanced evolutionary states. 
It is demonstrated that less evolved clusters have shallower mass profiles and their galaxy orbits are 
more radially biased at the virial sphere. This property is consistent with the expected evolution of 
the mass profiles as well as with the observed orbital segregation of late and early type galaxies.

\end{abstract}


\section{Introduction}

Kinematic data for galaxy clusters offer a unique possibility to study their mass distribution and the orbits of 
the galaxies within them. Depending on the quality of data sample and the purpose of the analysis 
a number of methods of data modelling has been proposed, from the mass estimate based on the virial theorem 
\citep{Hei85,Gir98}, the Jeans analysis with the two even velocity moments \citep{Lok03} and the caustic technique 
\citep{Dia97} to the recently presented kinematic deprojection with the mass inversion \citep{Mam10} and the projected 
phase-space analysis with a fully anisotropic model of the distribution function 
\citep{Woj09}. There is no doubt that the component which is best constrained with all methods of data modelling 
is the mass profile. The analysis of galaxy kinematics in nearby galaxy clusters revealed that the mass distribution 
is consistent with the universal NFW \citep{Nav97} profile \citep[see e.g.][]{Biv03,Lok06,Rin06}, which was confirmed in other 
studies based on X-ray observations \citep[see e.g.][]{Poi05} or lensing data \citep[e.g.][]{Com06}. On the other hand, 
constraints on the parameters of the universal density profile of individual clusters from galaxy kinematics are available 
for a very limited number of objects \citep[see][]{Lok06,Woj07}. One of the main aims of this work is to provide robust 
estimates of these parameters for a statistically significant sample of nearby ($z<0.1$) galaxy clusters.

Constraining the anisotropy of galaxy orbits from kinematic data is a difficult task. The main problem 
arises from the so-called mass-anisotropy degeneracy which occurs in the Jeans analysis 
of the velocity dispersion profile \citep{Mer87}. In order to break this degeneracy a few solutions have 
been proposed. \citet{Bin82} derived the velocity anisotropy profile by combining the Jeans standard analysis with 
an independent constraint on the mass profile. This approach was adopted by several authors who used the estimate of the mass profile 
from the kinematics of ellipticals whose orbits appear to be distributed isotropically \citep{Biv04}, lensing observations 
\citep[][]{Nat96} or X-ray data \citep[][]{Ben06,Hwa08}. Another method to solve the problem of the mass-anisotropy 
degeneracy relies on modelling the degree of non-gaussianity in the velocity distribution \citep[][]{Mer90}. 
A way to account for this effect is to use two even velocity moments \citep[][]{Lok03,Lok06}. Apart from this, 
in a few attempts a full model of the distribution function has been used, but in all cases a flat anisotropy 
profile was assumed \citep[see e.g.][]{Mar00,Mah04}. Nevertheless, this method is still promising. As shown 
by \cite{Woj09} in a series of tests on mock kinematic data, the projected phase-space analysis with a suitable model 
of the distribution function proposed by \citet{Woj08} allows to place robust constraints on the spatial variation 
of the orbital anisotropy within the virial sphere as well as on the total mass profile. In this paper, we follow the 
method described in \citet{Woj09} to constrain the orbital anisotropy at the cluster centres and around 
the virial radius.

Most measurements of the orbital anisotropy in galaxy clusters have been obtained under the assumption of a flat profile. All results are 
consistent within errors with isotropic orbits \citep[see e.g.][]{Mar00,Rin03,Mah04,Lok06,Woj07}. Several studies were devoted 
to constraining the spatial variation of the orbital anisotropy. The analysis of galaxy kinematics of individual clusters points 
to rather isotropic orbits, although the anisotropy exhibits a significant scatter between clusters \citep[see e.g.][]{Nat96,Ben06,Hwa08}. 
More robust constraints on the anisotropy profile were obtained in the analysis of the composite clusters resulting from combining a number 
of kinematic data sets into one. It was demonstrated that the orbits of early type galaxies tend to be isotropic within the virial sphere, 
whereas late type galaxies are on predominantly radial orbits at the virial radius \citep{Ada98,Biv04}. This property is referred 
to as the orbital segregation and is commonly attributed to the orbital evolution from radially dominated orbits for spirals to isotropic 
ones for ellipticals. We note, however, that the orbital structure of both morphological types of galaxies can hardly be reconciled with 
the result of global isotropy. This problem is also addressed in this paper.

The paper is organized as follows. In section 2, we describe our cluster sample as well as kinematic data and the 
method of interloper removal. In section 3, we outline the details of our method of the projected phase-space analysis. 
The results of the analysis are presented in section 4. In the same section, we discuss the constraints on the 
mass-concentration relation and study the mass scaling relations. Section 5 is the devoted to comparison between our 
constraints on the parameters of the mass profile and those found in the literature and based on other methods. 
In section 6, we present two tests of consistency between the data and some assumptions underlying our analysis. 
The summary follows in section 7. In this work, we adopted a flat $\Lambda$CDM cosmology with density parameter 
$\Omega_{\rm m}=0.3$ and the Hubble constant $H_{0}=70$ km s$^{-1}$ Mpc$^{-1}$.

\section{Data}

\subsection{Cluster selection}
The primary criterion for cluster selection is the number of galaxies with spectroscopic redshifts per cluster. 
An optimum choice of a threshold for this number should result from the consideration of two contradictory effects. 
On one hand, the quality of statistical reasoning requires the number of redshifts to be as large as possible, 
but on the other we wish to build a sample with a statistically significant number of clusters which means that 
we also have to include clusters with a rather low number of spectroscopic redshifts. Trying to find a compromise 
we checked the number of available redshifts per cluster, as provided by NASA/IPAC Extragalactic 
Database (NED). We found that the optimal choice for the purpose of our analysis is to consider clusters with 
at least $70$ galaxies within the aperture of $2.5$ Mpc and with velocities from the range $\pm 4000$ km s$^{-1}$ 
in the rest frame of the cluster. The radius of the aperture equals to the virial radius of massive galaxy clusters 
and its role is to truncate all clusters to the approximate size of the virial sphere (in projection). 
%To be more precise, it corresponds 
%to the virial mass of $9\times 10^{14}M_{\odot}$. 
The velocity range $\pm4000$ km s$^{-1}$ is a commonly used velocity cut-off 
which allows to perform a preliminary separation of cluster members from the interlopers 
(galaxies of background or foreground).

The second criterion for cluster selection concerns their dynamical state. The reason for considering this property 
is the fact that the basic assumption underlying our method of dynamical analysis is the state of equilibrium. It means 
that our sample should be possibly free of dynamically disturbed clusters, especially major mergers. In order to identify 
such clusters we studied X-ray images of clusters obtained with ASCA \citep{Hor01}, XMM-Newton \citep{Sno08} 
and ROSAT \citep{Sch01} satellites. We rejected all clusters whose images exhibit signatures of merger activity in the form of clearly 
visible distortions (e.g. A3667) or the presence of massive subclusters (e.g. A3558, the most massive cluster in the 
Shapley Concentration).

Another approach to assess the dynamical state of clusters was to study the general properties of galaxy kinematics. 
This method was particularly important in the case of clusters for which X-ray data were not available. Following this approach, we 
excluded all clusters with clearly asymmetric or bimodal velocity distributions which are good signatures of a merger or at least 
an accidental alignment of some structures. Let us note that the analysis of galaxy kinematics was carried out after the proper 
removal of interlopers (see subsection 2.4).

As an independent test, we also checked for the presence of a cool core which appears to correlate with the dynamical state of clusters 
\citep[][]{Bur08,San09,Hud10} and may be regarded as an indicator of how relaxed the cluster is. We note, however, that the 
presence of a cool core was a secondary criterion of our selection. We used this criterion only in cases when the conclusions drawn 
from the kinematics or the X-ray image seemed to be ambiguous or contradictory. 

Finally, we also excluded clusters which appear to be accompanied by other clusters at comparable redshifts and relative positions on the sky 
smaller than $5$ Mpc which is around twice the expected virial radius for a typical cluster. The majority of them are binary clusters at the 
initial state of merging, e.g. A399 and A401. The main reason for rejecting these systems is the fact that their velocity diagrams are 
severely contaminated by galaxies from neighbouring clusters. Due to the proximity of these clusters both on the sky and in redshift space, 
velocity diagrams overlap and the selection of cluster members becomes impossible or at least highly ambiguous.

\subsection{Cluster sample}
In order to find all nearby ($z<0.1$) galaxy clusters which satisfy selection criteria described above we searched 
NED and the recently published, but not yet included in NED, spectroscopic data from WINGS (The WIde field Nearby Galaxy cluster Survey) 
which is a photometric and spectroscopic survey of a few tens of nearby galaxy clusters \citep{Cav09}. The final sample resulting from our 
data mining consists of $41$ galaxy clusters, out of which $40$ are rich clusters from the Abell catalogue \cite[][]{Abe58,Abe89} and 
one (MKW 4) is a poor cluster containing a cD galaxy \citep[][]{Mor75}. Table \ref{tab_prop} provides the list of all 
clusters as well as their basic observational properties, such as the heliocentric redshift $z$, the number of spectroscopic redshifts 
within the aperture of $2.5$ Mpc and the velocity range $\pm 4000$ km s$^{-1}$ around the mean $N_{\rm tot}$ and after the removal of 
interlopers (see subsection 2.4) $N_{\rm mem}$.

\begin{table}
\begin{center}
\begin{tabular}{lcrrrl}
Name   &  $z$    & $N_{\rm tot}$ & $N_{\rm mem}$ & B-M type & CC \\
\hline
A0085 &  0.0551 & 325  & 317 & I       & yes$^{\rm a}$ \\
A0119 &  0.0442 & 215 & 211 & II-III & no$^{\rm a}$\\
A0133 &  0.0566 & 75  & 75  & I-II   & yes$^{\rm a}$\\
A0262 &  0.0163 & 141  & 135 & III     & yes$^{\rm a}$ \\
A0376 &  0.0484 & 85   & 85  & I-II    & ? \\
A0496 &  0.0329 & 324  & 313 & I       & yes$^{\rm a}$ \\
A0539 &  0.0284 & 141  & 126 & III     & yes$^{\rm b}$ \\
A0576 &  0.0389 & 241  & 214 & III     & yes$^{\rm a}$ \\
A0671 &  0.0502 & 129  & 124 & II-III  & no$^{\rm d}$ \\
A0779 &  0.0225 & 154  & 139 & I-II    & yes$^{\rm d}$ \\
A0954 &  0.0932 & 68   & 66  & I-II    & ? \\
A0957 &  0.0436 & 138  & 124 & I-II    & no$^{\rm d}$ \\
A0978 &  0.0544 & 100  & 90  & II      & ? \\
A1060 &  0.0126 & 369 & 365 & III    & yes$^{\rm a}$\\
A1139 &  0.0398 & 147 & 143 & III    & ? \\
A1190 &  0.0751 & 137 & 122 & II     & ? \\
A1314 &  0.0335 & 140 & 127 & III    & no$^{\rm d}$\\
A1650 &  0.0838 & 212 & 204 & I-II   & yes$^{\rm a}$\\
A1691 &  0.0721 & 118 & 110 & II     & ? \\
A1767 &  0.0703 & 152  & 147 & II      & no$^{\rm d}$ \\
A1773 &  0.0765 & 109  & 103 & III     & ? \\
A1795 &  0.0625 & 177  & 163 & I       & yes$^{\rm a}$\\
A1809 &  0.0791 & 131  & 123 & II      & no$^{\rm d}$ \\
A1983 &  0.0436 & 117  & 100 & III     & yes$^{\rm d}$ \\
A2052 &  0.0355 & 150  & 107 & I-II    & yes$^{\rm a}$ \\
A2063 &  0.0349 & 150  & 115 & II      & yes$^{\rm a}$ \\
A2107 &  0.0411 & 88   & 85  & I       & yes$^{\rm c}$ \\
A2142 &  0.0909 & 229  & 226 & II      & yes$^{\rm a}$ \\
A2175 &  0.0951 & 93   & 87  & II      & ? \\
A2415$^w$ &  0.0581 & 98   & 96  & III     & yes$^{\rm d}$ \\
A2593 &  0.0413 & 230  & 210 & II      & no$^{\rm d}$ \\
A2634 &  0.0314 & 223  & 185 & II      & yes$^{\rm a}$ \\
A2670 &  0.0762 & 246  & 238 & I-II    & yes$^{\rm c}$ \\
A2734 &  0.0625 & 174  & 149 & III     & no$^{\rm b}$ \\
A3158 &  0.0597 & 148  & 145 & I-II    & no$^{\rm a}$ \\
A3571 &  0.0391 & 180  & 168 & I       & yes$^{\rm a}$ \\
A3581 &  0.0230 & 82   & 74  & I       & yes$^{\rm a}$ \\
A3809$^w$ &  0.0623 & 105 & 99 & III     & ? \\
A4059 &  0.0475 & 237  & 192 & I       & yes$^{\rm a}$ \\
AS805 &  0.0139 & 159  & 148 & I       & ? \\
MKW4 &  0.0200  & 132  &124  & I       & yes$^{\rm a}$ \\
\end{tabular}
\caption{Observational properties of galaxy clusters selected for the analysis. The table includes the cluster name, 
heliocentric redshift ($z$), number of galaxy redshifts within the aperture $2.5$ Mpc and the velocity range $\pm 4000$km s$^{-1}$ ($N_{\rm tot}$), 
number of redshifts after interloper removal ($N_{\rm mem}$), Bautz-Morgan morphological type (B-M type) and the comment on a cool core 
judged on the basis of: $^{\rm a}$\citet{Hud10}, $^{\rm b}$\citet{Che07}, $^{\rm c}$\citet{Whi00}, $^{\rm d}$\citet{Whi97}. Superscript 
$w$ marks clusters from the WINGS survey.}
\label{tab_prop}
\end{center}
\end{table}

All clusters in Table~\ref{tab_prop} are relaxed objects in terms of merger activity.
However, we cannot expect that the whole cluster sample is completely uniform in terms of the evolutionary stage. In order to check this, 
we searched for two indirect indicators of cluster evolution. The first indicator is the presence of a cool core. As shown by \citet{Bur08} 
and \citet{Hud10}, the survival of a cool core in nearby galaxy clusters is likely associated with the quiescent phase of recent history, 
when the cluster has not experienced major mergers. Therefore the clusters which possess cool cores are expected to be more relaxed. 
The second indicator is the morphological type. We used the Bautz-Morgan classification \citep[][]{Bau70} which describes the degree 
to which a cluster is dominated by the BCG (bright central galaxy). 
%The early type (I) of Bautz-Morgan classification is assigned to clusters with a prominent cD (central diffuse) 
%galaxy in their centres, whereas the late type (III) is associated with clusters without a distinct BCG. Type II is the intermediate type 
%in this classification. 
Like for many other morphological classifications of galaxy clusters, the Bautz-Morgan morphological type appears to correlate with the dynamical 
state of cluster evolution \citep[][]{Sar88}. The early type clusters (I,I-II) which are more dominated by BCGs are more dynamically 
relaxed than clusters of late types (II-III,III).

The last two columns of Table \ref{tab_prop} include information on the presence of a cool core and the Bautz-Morgan morphological type. 
Observational constraints on a cool core were obtained from the literature \citep[][]{Whi97,Whi00,Che07,Hud10}. We note that the criteria used as 
the signatures of a cool core were not common to all authors and were mainly based on the inspection of such effects as 
a temperature drop at small radii, substantial mass deposition rate or sufficiently small cooling time in the cluster centres. 
%(cool cores manifest themselves as a drop in the temperature profile at small radii) 
%(cool core clusters appear to exhibit cooling times smaller than the Hubble time). 
Due to the fact that our cluster sample is optically selected, the X-ray data are not equally accessible for all objects. In particular, 
the information on the presence of a cool core was found for only $31$ clusters ($22$ cool core clusters and $9$ non-cool core clusters). 
For other clusters X-ray data are very limited and therefore insufficient to study this property.

The Bautz-Morgan morphological types of the clusters were obtained from NED. We note that this classification is not objective and may be 
treated as indicative only. Nevertheless, it allows us to divide our cluster sample into two classes, the early and late 
type clusters, which are thought to represent more and less advanced dynamical state of cluster evolution. A similar division of the clusters 
may be also carried out with respect to the presence of a cool core. In this approach cool core clusters are expected to be 
more dynamically evolved objects. A closer inspection of Table~\ref{tab_prop} shows that both classifications are clearly related 
to each other: the early type (I,I-II) clusters constitute $60$ per cent of the cool core clusters and just $20$ per cent of the non-cool 
core clusters. Let us note that this correlation, although not very tight, increases our confidence in using both classifications to assess 
the dynamical state of a cluster.

\subsection{Velocity diagrams}

The positions and redshifts of galaxies in the field of the clusters from Table~\ref{tab_prop} were obtained from NED. 
This database provides the most up-to-date compilation of many different kinds of extragalactic data, including spectroscopic 
redshifts. The majority of spectroscopic data stored in NED comes from data releases of large scale redshift surveys, e.g. the Sloan Digital 
Sky Survey \cite[see][for the latest data release]{Ade08} or the 2dF Galaxy Redshift Survey \citep{Col01}, as well as a number of 
spectroscopic surveys dedicated to observations of individual galaxy clusters, e.g. the NOAO Fundamental Plane Survey \citep{Smi04}. 
We note that the SDSS, the 2dF survey and the NFPS are the three dominant sources of redshift data for our cluster sample. For two clusters, 
A2415 and A3809, redshifts and positions come from the WINGS survey \citep[][]{Cav09}. We found that in both cases the number of 
redshifts provided by the catalogue of this survey is considerably higher than that provided by NED. 

In order to calculate the projected clustercentric distances $R$ of all galaxies one needs to find cluster centres. From the observational point of view, 
the cluster centre may be defined as the position of the X-ray emission peak or the location of the BCG. For clusters that are not dynamically 
disturbed both centres coincide with each other and both point very well to the minimum of the gravitational potential well. 
Since our cluster sample was constructed using optical data, we decided to assign cluster 
centres as the BCG positions. We found that $39$ clusters from the sample have a well-confirmed central galaxy classified in 
NED as a cD galaxy ($27$ clusters) or a BCG ($12$ clusters). For two clusters, A2415 and A3809, central bright galaxies were found 
by visual inspection of optical images. Both clusters have distinct, bright and extended ellipticals lying in the close vicinity of 
the cluster centre given by the catalogue of Abell clusters.

With cluster centres given by the positions of BCGs, we calculated the projected clustercentric distances of all galaxies. Except of a few 
clusters, the distances extend to $R_{\rm max}=2.5$ Mpc which is a fixed size of the aperture in our analysis. 
Three exceptional clusters are A376, A2415 and A3809. The velocity diagram of the first of these is limited to the aperture of around $1.8$ Mpc 
which may be the effect of the incompleteness of the spectroscopic survey or may reflect extreme compactness of this cluster. Galaxy distances 
of the other two clusters are limited by the aperture (around $1.5$ Mpc) to which the WINGS survey is complete \citep[see][]{Cav09}. 
For the analysis of these three clusters we assumed $R_{\rm max}=1.5$ Mpc. We note that an independent test for data completeness is 
presented in subsection 6.1.

\subsection{Interloper removal}

In order to separate cluster galaxies from the interlopers, i.e. galaxies of background or foreground, we applied the method 
proposed by \citet{Har96} which appears to be one of the most effective algorithms for interloper removal \citep[see][]{Woj07a,Woj07}. 
In this approach the interlopers are identified as galaxies with velocity along the line of sight exceeding a maximum velocity evaluated 
for a given projected radius $R$. Calculation of the maximum velocity $v_{\rm max}(R)$ is based on a toy model of cluster dynamics which 
assumes that galaxies follow circular orbits with velocity $v_{\rm cir}=\sqrt{GM(r)/r}$ or fall towards the cluster centre with velocity 
$\sqrt{2}v_{\rm cir}$. Evaluation of $v_{\rm max}(R)$ in such a model is simplified to finding the maximum of both velocity vectors projected 
onto the line of sight. We note that the interloper removal is an iterative procedure, i.e. the scheme is repeated several times until 
convergence is achieved. The mass $M(r)$ in each iteration is approximated by the virial mass estimator \citep[][]{Hei85} evaluated 
for all galaxies within the aperture $R=r$.

\section{Overview of the method}

In this section, we summarize our method of statistical inference of the mass and anisotropy profiles from galaxy kinematics, 
as developed and described in detail in \citet{Woj09}. The method relies on the Markov Chain Monte Carlo (MCMC) analysis of galaxy 
distribution in the projected phase-space, i.e. the space of the projected radii $R$ and velocities along the line of sight $v_{\rm los}$, 
in terms of a properly parametrized model of the distribution function. The projected phase-space density $f_{\rm los}(R,v_{\rm los})$ is 
related to the distribution function $f(E,L)$ through the following integral
\begin{equation}\label{f_los_def}
	f_{\rm los}(R,v_{\rm los})=2\pi R\int_{-z_{\rm max}}^{z_{\rm max}}\!\!\!\!\textrm{d}z\int\!\!\!\int_{E>0}\!\!\!\!
	\textrm{d}v_{R}\textrm{d}v_{\phi}f(E,L),
\end{equation}
where $z$ is the distance along the line sight from the cluster centre, $v_{R}$ and $v_{\phi}$ are velocity components in cylindrical 
coordinates, $E$ is the positively defined binding energy per unit mass, $L$ is the absolute value of the specific angular momentum 
and $z_{\rm max}$ is the distance of equality between $v_{\rm los}$ and the escape velocity for a fixed $R$. We note that the 
assumption underlying the functional form of both $f_{\rm los}(R,v_{\rm los})$ and $f(E,L)$ is spherical symmetry and the state of 
equilibrium.

In our analysis we adopted the model of the distribution function proposed by \citet{Woj08}. The model was adjusted to the phase-space 
properties of dark matter particles in cluster-size haloes formed in cosmological $N$-body simulations. It assumes separability in the 
energy and the angular momentum, i.e. $f(E,L)=f_{E}(E)f_{L}(L)$, in which the angular momentum part is expressed by the following analytical 
ansatz
\begin{equation}	\label{f_L}
	f(L)=\Big(1+\frac{L^{2}}{2L_{0}^{2}}\Big)^{-\beta_{\infty}+\beta_{0}}
	L^{-2\beta_{0}}.
\end{equation}
$\beta_{0}$ and $\beta_{\infty}$ in (\ref{f_L}) are two asymptotic values of the anisotropy parameter
\begin{equation}
\beta(r)=1-\frac{\sigma_{\theta}^{2}(r)}{\sigma_{r}^{2}(r)}
\end{equation}
describing a ratio of the radial $\sigma_{r}$ to tangential $\sigma_{\theta}$ velocity dispersion. The profile of the anisotropy is 
a monotonic function changing from $\beta_{0}$ at the cluster centre to $\beta_{\infty}$ for large radii \citep[see][]{Woj08}. The 
scale of transition between these two asymptotes is specified by parameter $L_{0}$.

The energy part $f_{E}(E)$ is given by the integral equation defining the relation between the distribution function and the 
assumed density profile of a tracer $\rho(r)$
\begin{equation}\label{rho_DF}
	\rho(r)=\int\!\!\!\int\!\!\!\int f_{E}(E)
	\Big(1+\frac{L^{2}}{2L_{0}^{2}}\Big)^{-\beta_{\infty}+\beta_{0}}
	L^{-2\beta_{0}}\textrm{d}^{3}v.
\end{equation}
This equation can be simplified to the form of a one-dimensional integral and solved numerically for $f_{E}(E)$ 
(see Appendix B in \citet{Woj08}). In the following, we approximate the total density distribution by the universal 
NFW profile \citep{Nav97}, i.e.
\begin{equation}\label{NFW}
	\rho(r/r_{\rm s})=\frac{1}{4\pi(\ln2-1/2)} \frac{1}{(r/r_{\rm s})(1+r/r_{\rm s})^{2}}
	\frac{M_{\rm s}}{r_{\rm s}^{3}},
\end{equation}
where $r_{\rm s}$ is the scale radius and $M_{\rm s}$ is the mass enclosed in a sphere of this radius. We also assume that 
galaxies trace dark matter. This assumption, commonly referred to as a hypothesis of constant mass-to-number density 
profile, was shown to be fairly held for relatively large clustercentric radii \citep[see e.g.][]{Car97,Biv03}. 
On the other hand, its robustness is more uncertain in the cluster centre: observations suggest that the profile of galaxy distribution 
at small radii may be as steep as dark matter \citep[][]{Lin04} or exhibit a core \citep[][]{Pop07}. The problem of the 
consistency between the assumption about mass-to-number ratio and the data collected for our analysis is addressed in 
section 6.
%We address the problem of the robustness of the assumption in section 6.
%Nevertheless, we address the problem of the robustness of this assumption in section 6.

Although $M_{\rm s}$ and $r_{\rm s}$ set the most natural scales for the phase-space units and both are used as the primary 
parameters of the mass profile in our analysis, a common way to parametrize the universal mass profile is to use the 
virial mass $M_{\rm v}$ and the concentration parameter $c$ defined by equations
\begin{eqnarray}
	\frac{M_{\rm v}}{(4/3)\pi r_{\rm v}^{3}} & = & \Delta_{\rm c}\rho_{\rm c} \\ \nonumber
	c & = & r_{\rm v}/r_{\rm s},
\end{eqnarray}
where $r_{\rm v}$ is the virial radius, $\rho_{\rm c}$ is the present critical density and $\Delta_{\rm c}$ is the virial 
overdensity. For the cosmological model assumed in this paper $\Delta_{\rm c}=102$ \citep{Bry98,Lok01a}.

Constraints on the parameters of the mass and anisotropy profiles were determined by means of the analysis of the posterior 
probability with the MCMC technique. Following \citet{Woj09} we used the likelihood function defined by
\begin{equation}
	L\propto \prod_{i=1}^{N_{\rm mem}} f_{\rm los}(R_{i},v_{{\rm los},i}|M_{\rm s},r_{\rm s},\beta_{0},\beta_{\infty}),
\end{equation}
where $i$ is the reference number of a galaxy. The projected phase-space density (\ref{f_los_def}) was evaluated 
numerically using the algorithm of Gaussian quadrature, as outlined in \citet{Woj09}. We note that $f_{\rm los}$ 
preserves the normalization within the aperture $R_{\rm max}$ fixed in subsection 2.3. For the scale parameters 
we used the Jeffreys priors, i.e. $p(r_{\rm s})\propto 1/r_{\rm s}$ and $p(M_{\rm s})\propto 1/M_{\rm s}$. In 
order to assign equal weights for all types of the orbital anisotropy we adopted 
$\beta\rightarrow(-1/2)\ln(1-\beta)=\ln(\sigma_{r}/\sigma_{\theta})$ reparametrization and assumed flat priors 
for redefined parameters of the anisotropy profile. We fixed the value of $L_{0}$ parameter at 
$0.45r_{\rm s}\sqrt{GM_{\rm s}/r_{\rm s}}$ which corresponds to $\sim1r_{\rm s}$ scale of the transition between $\beta_{0}$ 
and $\beta_{\infty}$, as found for massive dark matter haloes from cosmological simulations \citep[see][]{Woj08}. 
In order to keep positivity of the phase-space density in the cluster centre we narrowed the range of the prior 
for the central anisotropy to $\beta_{0}\leq 1/2$ \citep{An06}. We also placed an upper limit on the outer 
anisotropy, $\beta_{\infty}=0.99$,  which solves the problem of an improper posterior distribution for models 
with $\beta_{\infty}\approx 1$.

We determined the Markov chains using the Metropolis-Hastings algorithm \citep[see][]{Gre05,Woj09}. All free 
parameters of the algorithm were properly adjusted to keep the acceptance rate (the probability of parameter change 
between each two neighbouring models of the Markov chain) at the level of $\sim30$ per cent which is the recommended 
value for many-parameters models \citep[][]{Gel04}. Our Markov chains consist of $2\times 10^{4}$ models. We checked 
that this length is sufficient to properly explore our parameter space, also in a joint modelling described 
in subsection 4.5. The convergence of the MCMC algorithm has been tested by means of monitoring the profile of the posterior 
probability along the chains, the analysis of the autocorrelation functions calculated for all parameters in use and 
the comparison of the parameter dispersions evaluated in different sections of the Markov chains. In this approach, properly 
mixed Markov chain is characterized by the posterior probability fluctuating around a flat profile, fast decay of the 
autocorrelation functions and possibly small variations of the parameter dispersions.

\section{Results of the analysis}

\begin{figure*}
\begin{center}
    \leavevmode
    \epsfxsize=12cm
    \epsfbox[50 50 1200 1200]{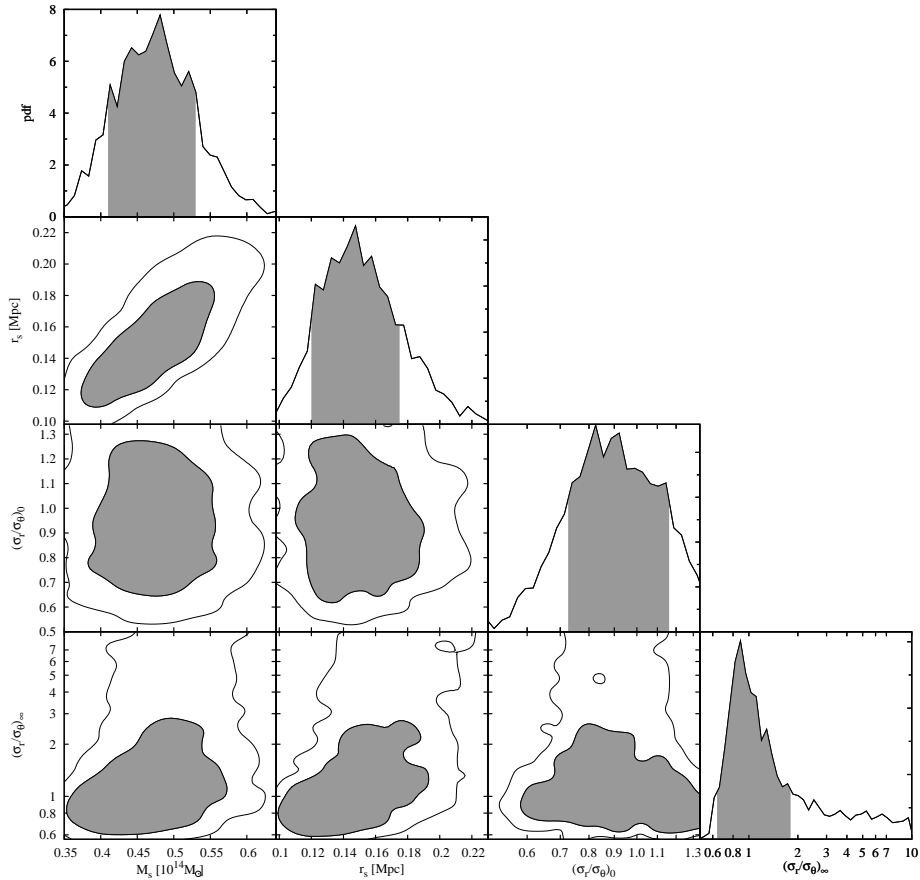}
\end{center}
\caption{Results of the MCMC analysis of the velocity diagram for the A1060 galaxy cluster. Black solid lines are the contours of 
the $1\sigma$ and $2\sigma$ credibility areas (panels below the diagonal) or the profiles of the marginal probability distribution 
(panels along the diagonal). Gray areas indicate the $1\sigma$ credibility ranges of the parameters.}
\label{A1060}
\end{figure*}

We analyzed velocity diagrams of $41$ nearby galaxy clusters in terms of the projected phase-space density, as described in section 3. 
Fig.~\ref{A1060} shows an example of the results obtained for the richest velocity diagram of the A1060 galaxy cluster. The contours 
show the $1\sigma$ and $2\sigma$ credibility constraints on the parameters of the mass profile, $M_{\rm s}$ and $r_{\rm s}$, and the 
profile of the orbital anisotropy, $\beta_{0}$ and $\beta_{\infty}$. The panels along the diagonal and in the top right corner show 
the profiles of the marginal probability distribution and indicate the $1\sigma$ credibility range for all parameters in use. We note 
that the $1\sigma$ and $2\sigma$ credibility ranges or contours are defined as those containing respectively $68$ or $95$ per cent of 
the corresponding marginal probability.

\begin{table*}
 \begin{footnotesize}
\begin{center}
\begin{tabular}{l|cc|cc|cc|cc}
cluster& $M_{\rm s}[10^{14}M_{\odot}]$ & $r_{\rm s}$[Mpc] & $M_{\rm v}[10^{14}M_{\odot}]$ & $c$ & $(\sigma_{r}/\sigma_{\theta})_{0}$ & 
$(\sigma_{r}/\sigma_{\theta})_{r_{\rm v}}$ & $\beta_{0}$ & $\beta_{r_{\rm v}}$ \\
\hline
  A0085&  $  2.81_{- 0.30}^{+ 0.53}$&  $  0.65_{- 0.14}^{+ 0.16}$&  $ 12.33_{- 1.34}^{+ 1.78}$&  $  4.25_{- 0.96}^{+ 0.76}$
&  $  1.16_{- 0.12}^{+ 0.16}$&  $  2.04_{- 0.25}^{+ 0.12}$
&  $  0.26_{- 0.18}^{+ 0.17}$&  $  0.76_{- 0.07}^{+ 0.03}$  \\
  A0119&  $  1.08_{- 0.12}^{+ 0.24}$&  $  0.45_{- 0.09}^{+ 0.11}$&  $  5.05_{- 0.75}^{+ 0.73}$&  $  4.61_{- 1.24}^{+ 0.79}$
&  $  1.16_{- 0.13}^{+ 0.17}$&  $  2.06_{- 0.24}^{+ 0.13}$
&  $  0.26_{- 0.20}^{+ 0.18}$&  $  0.76_{- 0.07}^{+ 0.03}$  \\
  A0133&  $  0.69_{- 0.13}^{+ 0.28}$&  $  0.19_{- 0.05}^{+ 0.10}$&  $  5.69_{- 1.11}^{+ 1.47}$&  $ 11.27_{- 5.68}^{+ 1.49}$
&  $  1.22_{- 0.43}^{+ 0.13}$&  $  0.94_{- 0.25}^{+ 1.36}$
&  $  0.33_{- 0.93}^{+ 0.13}$&  $ -0.14_{- 0.95}^{+ 0.95}$  \\
  A0262&  $  0.32_{- 0.05}^{+ 0.08}$&  $  0.16_{- 0.04}^{+ 0.04}$&  $  2.51_{- 0.25}^{+ 0.43}$&  $ 10.47_{- 2.89}^{+ 2.52}$
&  $  1.02_{- 0.50}^{+ 0.09}$&  $  0.55_{- 0.15}^{+ 0.21}$
&  $  0.04_{- 2.68}^{+ 0.14}$&  $ -2.26_{- 2.76}^{+ 1.54}$  \\
  A0376&  $  0.80_{- 0.12}^{+ 0.30}$&  $  0.25_{- 0.08}^{+ 0.08}$&  $  5.56_{- 0.91}^{+ 1.59}$&  $  8.35_{- 3.16}^{+ 1.91}$
&  $  0.71_{- 0.19}^{+ 0.28}$&  $  1.89_{- 0.45}^{+ 0.22}$
&  $ -1.00_{- 1.71}^{+ 0.97}$&  $  0.72_{- 0.20}^{+ 0.05}$  \\
  A0496&  $  0.99_{- 0.11}^{+ 0.15}$&  $  0.42_{- 0.07}^{+ 0.08}$&  $  4.74_{- 0.58}^{+ 0.70}$&  $  4.78_{- 1.02}^{+ 0.63}$
&  $  0.70_{- 0.11}^{+ 0.17}$&  $  1.48_{- 0.32}^{+ 0.37}$
&  $ -1.04_{- 0.86}^{+ 0.72}$&  $  0.54_{- 0.29}^{+ 0.16}$  \\
  A0539&  $  0.42_{- 0.06}^{+ 0.09}$&  $  0.14_{- 0.03}^{+ 0.05}$&  $  3.75_{- 0.53}^{+ 0.57}$&  $ 12.89_{- 4.67}^{+ 1.66}$
&  $  0.92_{- 0.16}^{+ 0.24}$&  $  2.19_{- 0.26}^{+ 0.16}$
&  $ -0.18_{- 0.54}^{+ 0.44}$&  $  0.79_{- 0.06}^{+ 0.03}$  \\
  A0576&  $  1.71_{- 0.24}^{+ 0.29}$&  $  0.51_{- 0.10}^{+ 0.11}$&  $  8.18_{- 1.02}^{+ 1.31}$&  $  4.77_{- 1.24}^{+ 0.83}$
&  $  0.95_{- 0.15}^{+ 0.28}$&  $  1.91_{- 0.25}^{+ 0.21}$
&  $ -0.12_{- 0.48}^{+ 0.45}$&  $  0.72_{- 0.09}^{+ 0.05}$  \\
  A0671&  $  0.58_{- 0.08}^{+ 0.18}$&  $  0.14_{- 0.03}^{+ 0.06}$&  $  5.55_{- 0.79}^{+ 1.01}$&  $ 15.51_{- 6.06}^{+ 1.48}$
&  $  0.66_{- 0.10}^{+ 0.31}$&  $  1.24_{- 0.46}^{+ 0.92}$
&  $ -1.30_{- 0.86}^{+ 1.23}$&  $  0.35_{- 0.99}^{+ 0.44}$  \\
  A0779&  $  0.29_{- 0.04}^{+ 0.07}$&  $  0.24_{- 0.05}^{+ 0.09}$&  $  1.59_{- 0.30}^{+ 0.18}$&  $  5.84_{- 1.98}^{+ 1.08}$
&  $  1.14_{- 0.18}^{+ 0.17}$&  $  2.11_{- 0.39}^{+ 0.12}$
&  $  0.23_{- 0.32}^{+ 0.19}$&  $  0.78_{- 0.11}^{+ 0.02}$  \\
  A0954&  $  1.02_{- 0.29}^{+ 0.38}$&  $  0.39_{- 0.12}^{+ 0.15}$&  $  5.37_{- 1.18}^{+ 1.64}$&  $  5.43_{- 2.01}^{+ 1.61}$
&  $  0.53_{- 0.10}^{+ 0.23}$&  $  1.57_{- 0.47}^{+ 0.22}$
&  $ -2.58_{- 1.92}^{+ 1.83}$&  $  0.59_{- 0.42}^{+ 0.10}$  \\
  A0957&  $  0.40_{- 0.07}^{+ 0.11}$&  $  0.14_{- 0.03}^{+ 0.05}$&  $  3.58_{- 0.54}^{+ 0.54}$&  $ 13.01_{- 4.54}^{+ 1.75}$
&  $  0.81_{- 0.15}^{+ 0.37}$&  $  1.29_{- 0.41}^{+ 1.02}$
&  $ -0.51_{- 0.79}^{+ 0.79}$&  $  0.40_{- 0.67}^{+ 0.41}$  \\
  A0978&  $  0.96_{- 0.14}^{+ 0.32}$&  $  0.36_{- 0.11}^{+ 0.14}$&  $  5.23_{- 0.84}^{+ 1.35}$&  $  5.71_{- 2.43}^{+ 1.15}$
&  $  0.82_{- 0.24}^{+ 0.28}$&  $  1.87_{- 0.49}^{+ 0.22}$
&  $ -0.48_{- 1.45}^{+ 0.66}$&  $  0.71_{- 0.24}^{+ 0.06}$  \\
  A1060&  $  0.45_{- 0.04}^{+ 0.08}$&  $  0.14_{- 0.01}^{+ 0.04}$&  $  4.10_{- 0.36}^{+ 0.39}$&  $ 13.99_{- 3.31}^{+ 1.22}$
&  $  0.90_{- 0.18}^{+ 0.25}$&  $  0.88_{- 0.24}^{+ 0.77}$
&  $ -0.23_{- 0.72}^{+ 0.48}$&  $ -0.28_{- 1.16}^{+ 0.92}$  \\
  A1139&  $  0.41_{- 0.05}^{+ 0.12}$&  $  0.39_{- 0.08}^{+ 0.13}$&  $  1.57_{- 0.34}^{+ 0.24}$&  $  3.53_{- 1.25}^{+ 0.51}$
&  $  0.87_{- 0.19}^{+ 0.28}$&  $  1.75_{- 0.28}^{+ 0.22}$
&  $ -0.31_{- 0.83}^{+ 0.56}$&  $  0.67_{- 0.14}^{+ 0.07}$  \\
  A1190&  $  1.08_{- 0.19}^{+ 0.33}$&  $  0.50_{- 0.13}^{+ 0.17}$&  $  4.53_{- 1.07}^{+ 0.59}$&  $  4.00_{- 1.65}^{+ 0.60}$
&  $  1.30_{- 0.36}^{+ 0.10}$&  $  1.63_{- 0.16}^{+ 0.56}$
&  $  0.41_{- 0.52}^{+ 0.08}$&  $  0.62_{- 0.09}^{+ 0.17}$  \\
  A1314&  $  0.49_{- 0.07}^{+ 0.11}$&  $  0.21_{- 0.05}^{+ 0.06}$&  $  3.48_{- 0.47}^{+ 0.62}$&  $  8.64_{- 2.69}^{+ 1.44}$
&  $  0.87_{- 0.21}^{+ 0.27}$&  $  2.03_{- 0.44}^{+ 0.19}$
&  $ -0.33_{- 0.99}^{+ 0.55}$&  $  0.76_{- 0.15}^{+ 0.04}$  \\
  A1650&  $  2.52_{- 0.40}^{+ 0.69}$&  $  0.82_{- 0.14}^{+ 0.29}$&  $  8.19_{- 1.69}^{+ 1.48}$&  $  2.95_{- 1.08}^{+ 0.35}$
&  $  1.17_{- 0.18}^{+ 0.17}$&  $  1.56_{- 0.26}^{+ 0.48}$
&  $  0.27_{- 0.28}^{+ 0.17}$&  $  0.59_{- 0.18}^{+ 0.17}$  \\
  A1691&  $  1.38_{- 0.25}^{+ 0.54}$&  $  0.58_{- 0.17}^{+ 0.25}$&  $  5.36_{- 1.02}^{+ 1.69}$&  $  3.64_{- 1.58}^{+ 0.92}$
&  $  0.92_{- 0.17}^{+ 0.35}$&  $  1.77_{- 0.46}^{+ 0.29}$
&  $ -0.18_{- 0.60}^{+ 0.56}$&  $  0.68_{- 0.26}^{+ 0.08}$  \\
  A1767&  $  2.24_{- 0.46}^{+ 0.65}$&  $  0.50_{- 0.07}^{+ 0.26}$&  $ 11.84_{- 3.60}^{+ 1.40}$&  $  5.45_{- 2.51}^{+ 0.20}$
&  $  1.20_{- 0.46}^{+ 0.07}$&  $  0.78_{- 0.23}^{+ 1.12}$
&  $  0.30_{- 1.15}^{+ 0.07}$&  $ -0.66_{- 1.68}^{+ 1.39}$  \\
  A1773&  $  1.23_{- 0.26}^{+ 0.33}$&  $  0.42_{- 0.14}^{+ 0.13}$&  $  6.33_{- 1.10}^{+ 1.41}$&  $  5.24_{- 1.77}^{+ 1.59}$
&  $  1.11_{- 0.21}^{+ 0.25}$&  $  2.06_{- 0.34}^{+ 0.18}$
&  $  0.19_{- 0.42}^{+ 0.27}$&  $  0.76_{- 0.10}^{+ 0.04}$  \\
  A1795&  $  1.01_{- 0.21}^{+ 0.25}$&  $  0.23_{- 0.05}^{+ 0.09}$&  $  8.04_{- 1.10}^{+ 1.25}$&  $ 10.51_{- 3.64}^{+ 1.27}$
&  $  1.31_{- 0.35}^{+ 0.09}$&  $  0.51_{- 0.13}^{+ 0.23}$
&  $  0.42_{- 0.50}^{+ 0.07}$&  $ -2.79_{- 2.87}^{+ 2.00}$  \\
  A1809&  $  0.68_{- 0.14}^{+ 0.21}$&  $  0.22_{- 0.04}^{+ 0.10}$&  $  4.99_{- 0.65}^{+ 0.91}$&  $  9.20_{- 3.20}^{+ 0.97}$
&  $  0.91_{- 0.31}^{+ 0.35}$&  $  0.50_{- 0.12}^{+ 0.25}$
&  $ -0.21_{- 1.63}^{+ 0.59}$&  $ -2.94_{- 2.80}^{+ 2.16}$  \\
  A1983&  $  0.31_{- 0.05}^{+ 0.09}$&  $  0.29_{- 0.07}^{+ 0.12}$&  $  1.51_{- 0.33}^{+ 0.36}$&  $  4.79_{- 2.01}^{+ 0.94}$
&  $  1.08_{- 0.23}^{+ 0.32}$&  $  1.61_{- 0.23}^{+ 0.63}$
&  $  0.14_{- 0.53}^{+ 0.35}$&  $  0.61_{- 0.14}^{+ 0.19}$  \\
  A2052&  $  0.33_{- 0.05}^{+ 0.09}$&  $  0.14_{- 0.04}^{+ 0.05}$&  $  2.87_{- 0.49}^{+ 0.49}$&  $ 12.37_{- 4.84}^{+ 1.94}$
&  $  1.06_{- 0.21}^{+ 0.21}$&  $  2.27_{- 0.52}^{+ 0.14}$
&  $  0.11_{- 0.49}^{+ 0.27}$&  $  0.81_{- 0.13}^{+ 0.02}$  \\
  A2063&  $  0.60_{- 0.08}^{+ 0.23}$&  $  0.13_{- 0.03}^{+ 0.05}$&  $  5.95_{- 0.77}^{+ 1.19}$&  $ 16.15_{- 5.72}^{+ 1.56}$
&  $  0.67_{- 0.17}^{+ 0.39}$&  $  0.60_{- 0.18}^{+ 0.29}$
&  $ -1.20_{- 1.79}^{+ 1.31}$&  $ -1.79_{- 2.86}^{+ 1.52}$  \\
  A2107&  $  0.29_{- 0.03}^{+ 0.11}$&  $  0.11_{- 0.03}^{+ 0.05}$&  $  2.77_{- 0.44}^{+ 0.63}$&  $ 15.53_{- 6.23}^{+ 2.10}$
&  $  1.15_{- 0.34}^{+ 0.15}$&  $  0.98_{- 0.22}^{+ 1.42}$
&  $  0.25_{- 0.78}^{+ 0.17}$&  $ -0.05_{- 0.71}^{+ 0.87}$  \\
  A2142&  $  4.93_{- 0.81}^{+ 1.46}$&  $  0.99_{- 0.21}^{+ 0.31}$&  $ 16.70_{- 2.81}^{+ 2.41}$&  $  3.10_{- 1.08}^{+ 0.42}$
&  $  1.07_{- 0.19}^{+ 0.17}$&  $  1.61_{- 0.34}^{+ 0.37}$
&  $  0.13_{- 0.41}^{+ 0.22}$&  $  0.61_{- 0.23}^{+ 0.13}$  \\
  A2175&  $  2.11_{- 0.43}^{+ 0.99}$&  $  0.77_{- 0.22}^{+ 0.39}$&  $  6.88_{- 1.82}^{+ 1.71}$&  $  2.96_{- 1.67}^{+ 0.47}$
&  $  1.03_{- 0.23}^{+ 0.24}$&  $  1.82_{- 0.32}^{+ 0.19}$
&  $  0.05_{- 0.63}^{+ 0.33}$&  $  0.70_{- 0.14}^{+ 0.05}$  \\
  A2415&  $  0.57_{- 0.13}^{+ 0.19}$&  $  0.25_{- 0.08}^{+ 0.08}$&  $  3.66_{- 0.58}^{+ 0.90}$&  $  7.33_{- 2.58}^{+ 2.00}$
&  $  1.00_{- 0.24}^{+ 0.35}$&  $  2.08_{- 0.36}^{+ 0.25}$
&  $  0.01_{- 0.73}^{+ 0.45}$&  $  0.77_{- 0.11}^{+ 0.05}$  \\
  A2593&  $  0.55_{- 0.08}^{+ 0.09}$&  $  0.43_{- 0.09}^{+ 0.11}$&  $  2.08_{- 0.27}^{+ 0.37}$&  $  3.56_{- 1.05}^{+ 0.67}$
&  $  1.23_{- 0.13}^{+ 0.15}$&  $  2.03_{- 0.25}^{+ 0.11}$
&  $  0.34_{- 0.17}^{+ 0.13}$&  $  0.76_{- 0.07}^{+ 0.02}$  \\
  A2634&  $  0.77_{- 0.10}^{+ 0.17}$&  $  0.22_{- 0.03}^{+ 0.08}$&  $  5.83_{- 0.87}^{+ 0.73}$&  $  9.74_{- 2.68}^{+ 1.18}$
&  $  1.24_{- 0.38}^{+ 0.12}$&  $  0.79_{- 0.20}^{+ 0.47}$
&  $  0.35_{- 0.70}^{+ 0.11}$&  $ -0.59_{- 1.25}^{+ 0.96}$  \\
  A2670&  $  0.87_{- 0.06}^{+ 0.24}$&  $  0.16_{- 0.02}^{+ 0.06}$&  $  8.30_{- 0.90}^{+ 1.09}$&  $ 15.21_{- 4.80}^{+ 1.33}$
&  $  1.17_{- 0.33}^{+ 0.14}$&  $  0.82_{- 0.23}^{+ 0.92}$
&  $  0.27_{- 0.70}^{+ 0.15}$&  $ -0.49_{- 1.43}^{+ 1.16}$  \\
  A2734&  $  1.34_{- 0.26}^{+ 0.42}$&  $  0.57_{- 0.07}^{+ 0.29}$&  $  5.21_{- 1.87}^{+ 0.39}$&  $  3.65_{- 1.66}^{+ 0.14}$
&  $  0.66_{- 0.16}^{+ 0.20}$&  $  0.92_{- 0.29}^{+ 0.78}$
&  $ -1.28_{- 1.75}^{+ 0.93}$&  $ -0.19_{- 1.33}^{+ 0.84}$  \\
  A3158&  $  1.56_{- 0.17}^{+ 0.52}$&  $  0.25_{- 0.04}^{+ 0.10}$&  $ 12.93_{- 1.89}^{+ 2.45}$&  $ 11.47_{- 3.71}^{+ 1.41}$
&  $  0.83_{- 0.30}^{+ 0.27}$&  $  0.72_{- 0.34}^{+ 0.41}$
&  $ -0.44_{- 2.05}^{+ 0.61}$&  $ -0.92_{- 4.82}^{+ 1.14}$  \\
  A3571&  $  1.36_{- 0.24}^{+ 0.26}$&  $  0.25_{- 0.05}^{+ 0.07}$&  $ 10.83_{- 1.85}^{+ 1.20}$&  $ 10.62_{- 3.18}^{+ 1.92}$
&  $  1.09_{- 0.28}^{+ 0.18}$&  $  1.35_{- 0.35}^{+ 0.98}$
&  $  0.16_{- 0.66}^{+ 0.22}$&  $  0.45_{- 0.46}^{+ 0.36}$  \\
  A3581&  $  0.21_{- 0.02}^{+ 0.10}$&  $  0.12_{- 0.03}^{+ 0.07}$&  $  1.77_{- 0.31}^{+ 0.43}$&  $ 12.25_{- 6.07}^{+ 1.09}$
&  $  0.90_{- 0.15}^{+ 0.36}$&  $  1.14_{- 0.22}^{+ 1.18}$
&  $ -0.23_{- 0.52}^{+ 0.61}$&  $  0.23_{- 0.40}^{+ 0.58}$  \\
  A3809&  $  0.60_{- 0.08}^{+ 0.25}$&  $  0.45_{- 0.13}^{+ 0.17}$&  $  2.26_{- 0.50}^{+ 0.59}$&  $  3.52_{- 1.48}^{+ 0.77}$
&  $  0.92_{- 0.21}^{+ 0.35}$&  $  1.78_{- 0.34}^{+ 0.27}$
&  $ -0.19_{- 0.81}^{+ 0.57}$&  $  0.69_{- 0.17}^{+ 0.08}$  \\
  A4059&  $  1.16_{- 0.15}^{+ 0.23}$&  $  0.55_{- 0.09}^{+ 0.15}$&  $  4.45_{- 0.62}^{+ 0.63}$&  $  3.57_{- 0.96}^{+ 0.68}$
&  $  0.84_{- 0.12}^{+ 0.29}$&  $  1.72_{- 0.17}^{+ 0.24}$
&  $ -0.43_{- 0.53}^{+ 0.65}$&  $  0.66_{- 0.08}^{+ 0.08}$  \\
  AS805&  $  0.39_{- 0.05}^{+ 0.08}$&  $  0.19_{- 0.05}^{+ 0.05}$&  $  2.79_{- 0.40}^{+ 0.39}$&  $  8.86_{- 3.04}^{+ 1.42}$
&  $  1.05_{- 0.18}^{+ 0.21}$&  $  2.17_{- 0.35}^{+ 0.15}$
&  $  0.09_{- 0.44}^{+ 0.28}$&  $  0.79_{- 0.09}^{+ 0.03}$  \\
  MKW4&  $  0.23_{- 0.03}^{+ 0.05}$&  $  0.14_{- 0.03}^{+ 0.05}$&  $  1.81_{- 0.25}^{+ 0.27}$&  $ 10.43_{- 3.63}^{+ 1.38}$
&  $  1.06_{- 0.22}^{+ 0.20}$&  $  2.23_{- 0.36}^{+ 0.13}$
&  $  0.11_{- 0.51}^{+ 0.25}$&  $  0.80_{- 0.08}^{+ 0.02}$  \\
\end{tabular}
\caption{Constraints on the parameters of the mass and anisotropy profile of $41$ galaxy clusters from the MCMC analysis of their velocity 
diagrams. The table provides the MAP values and the ranges containing $68$ per cent of the corresponding marginal probability.}
\label{tab_para}
\end{center}
\end{footnotesize}
\end{table*}

To save space we do not show graphical representations of the results for other clusters. Instead, in Table~\ref{tab_para} we 
list numerical constraints on the parameters in the form of MAP (maximum a posteriori) values and the $1\sigma$ credibility ranges. 
Apart from the primary parameters of the mass profile used in our analysis, i.e. $M_{\rm s}$ and $r_{\rm s}$, Table~\ref{tab_para} 
also shows constraints on the virial mass $M_{\rm v}$ and the concentration parameter $c$. 
%For the assumed cosmological model 
%($\Omega_{m}=0.3$, $\Omega_{\Lambda}=0.7$, $h=0.7$) the virial overdensity parameter $\Delta_{c}$ defining the virial mass via equation 
%() equals $102$ \citep[see e.g.][]{Lok01a,Bry98}.

Since our model of the distribution function is well-defined only within the virial sphere, we decided to replace the parameter 
$\beta_{\infty}$ with $\beta_{r_{\rm v}}$ which is the anisotropy at the virial radius evaluated from the MAP value of the corresponding 
virial mass. The main motivation for this parameter replacement is to avoid the problem of the extrapolation of our model beyond the virial 
sphere. The new parameter of the orbital anisotropy was calculated using formulae for the velocity dispersions outlined in \citet{Woj09}.
The errors were evaluated by propagating the $1\sigma$ limits for the asymptotic values of the anisotropy profile to the virial radius. 
The final constraints on the anisotropy parameters are summarized in Table~\ref{tab_para} in terms of a commonly used anisotropy parameter, 
$\beta$, as well as in terms of the ratio of the radial-to-tangential velocity dispersion, $\sigma_{\rm r}/\sigma_{\theta}$.

\subsection{Scaling parameters of the mass profile}

\begin{figure}
\begin{center}
    \leavevmode
    \epsfxsize=8cm
    \epsfbox[50 50 580 420]{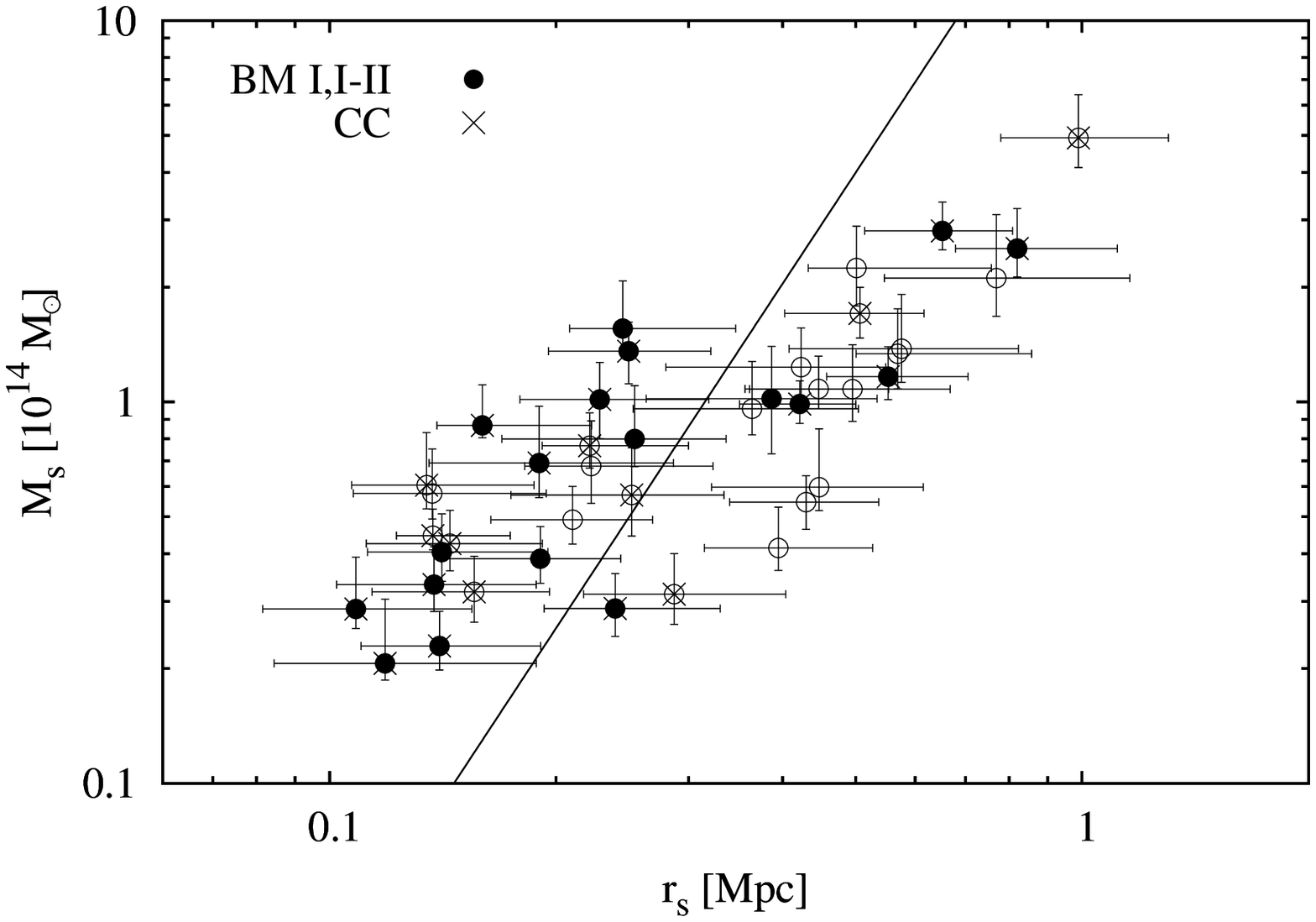}
\end{center}
\caption{Parameters of the mass profile obtained from the MCMC analysis of $41$ galaxy clusters, as listed in Table~\ref{tab_para}. 
Filled circles and crosses indicate early type (I,I-II type in Bautz-Morgan classification) clusters and cool core clusters. The solid 
line represents a family of density profiles with $c=7$ which is the median value of the concentration parameter in the cluster sample.}
\label{msrs_sca}
\end{figure}

\begin{figure}
\begin{center}
    \leavevmode
    \epsfxsize=8cm
    \epsfbox[50 50 580 420]{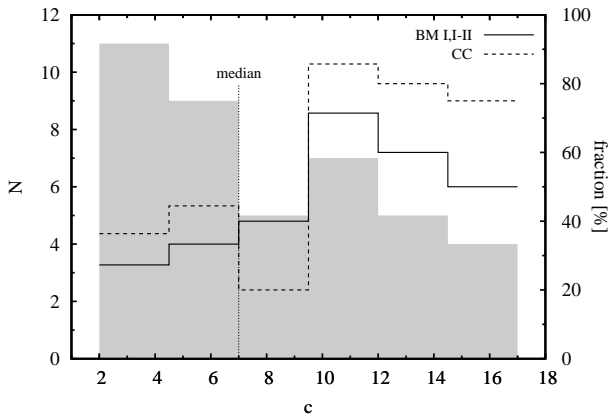}
\end{center}
\caption{The distribution of the concentration parameter from Table~\ref{tab_para} (shaded area). Dashed and solid lines show the relative 
numbers (right axis) of the cool core and early type (I, I-II type in Bautz-Morgan classification) clusters, respectively.}
\label{c_hist}
\end{figure}

Fig.~\ref{msrs_sca} shows the constraints on the scale radius $r_{\rm s}$ and the scale mass $M_{\rm s}$, as listed in Table~\ref{tab_para}. 
Both parameters exhibit a clear correlation which may be attributed mostly to the presence of a typical scale of the overdensity inside 
the sphere of the scale radius or, equivalently, to a typical value of the concentration parameter. In order to show this property 
more explicitly we plotted the line of constant $c$ equal to $7$ which is the median value of the concentration parameter for the 
clusters (see Fig.~\ref{c_hist}). This concentration parameter corresponds to the mean density inside the sphere of scale radius 
equal to $\sim 5600\rho_{c}$.

Parameters of individual clusters are scattered around the line of the mean overdensity. This scatter is also shown in the form 
of the distribution of the concentration parameter (see Fig.~\ref{c_hist}). Interestingly, clusters with well-confirmed 
cool cores as well as clusters of early morphological type tend to populate the left part of 
the diagram in Fig.~\ref{msrs_sca}. This means that the clusters which are dynamically more evolved exhibit higher central 
overdensity (or higher concentration parameter). This property is also well-seen in Fig.~\ref{c_hist}, where we plotted 
the relative number of both types of clusters in subsequent bins of the concentration parameter. Cool core clusters (early type 
clusters) contribute $70$ ($60$) per cent to the number of clusters with the concentration above the median value and $40$ ($30$) 
per cent to the number of clusters below the median value. Although the effect is not prominent, we can conclude that clusters 
with high concentrations are more likely to be dynamically more evolved than clusters with less concentrated mass distribution.

The correlation between the state of dynamical evolution of clusters and the steepness of the mass profile shown in 
Figs.~\ref{msrs_sca} and~\ref{c_hist} is in agreement with the results of recent cosmological simulations which point to 
two main factors responsible for this effect. First, more evolved clusters have not experienced severe mergers in their recent 
history and therefore tend to possess dark matter haloes with a steeper density profile \citep[see e.g.][]{Macc08}. Second, quiescent 
phase in the recent formation history of these clusters likely leads to the survival of cool cores \citep[see e.g.][]{Bur08,Hud10} 
and triggers the accumulation of baryonic gas due to the process of radiative cooling \citep[see e.g.][]{Gne04}. Higher concentration 
of gas in cluster centres deepens the gravitational potential well and consequently makes the density profile of dark matter steeper 
\cite[see e.g.][]{Dol09,Duf10}.

\subsection{The mass-concentration relation}

The concentration parameter and the virial mass are weakly anticorrelated \citep[][]{Nav97,Bul01}. This relation reflects the mass assembly 
histories of dark matter haloes and may be used as an independent test of cosmological models. Here we use our constraints on the virial 
mass and the concentration parameter to test theoretical mass-concentration relation against observations in the mass regime of galaxy 
clusters. 

\begin{figure}
\begin{center}
    \leavevmode
    \epsfxsize=8cm
    \epsfbox[50 50 580 420]{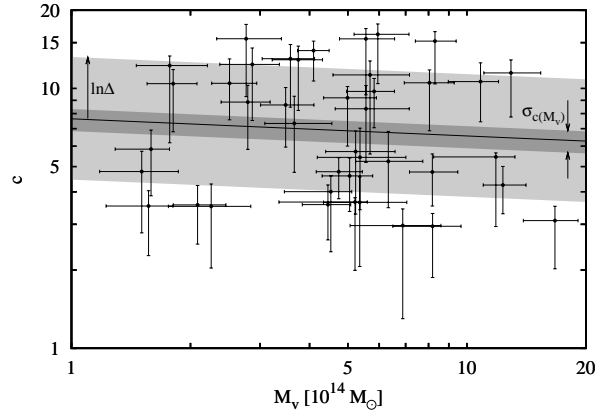}
\end{center}
\caption{The mass-concentration relation from the analysis of $41$ galaxy clusters. The solid line shows the best fit power-law profile 
for the mean concentration parameter. Shading indicates logarithmic dispersion of the concentration distribution, $\ln\Delta$, and 
the error of the normalization of the mean mass-concentration relation, $\sigma_{c(M_{\rm v})}$.}
\label{c_mv_dia}
\end{figure}

Fig.~\ref{c_mv_dia} shows our estimates of the virial mass $M_{\rm v}$ and the concentration parameter $c$. According to the results 
of cosmological simulations the scatter of the concentration parameter may be effectively modelled by a lognormal distribution 
\citep[][]{Bul01,Macc08}. The mean value of this distribution weakly varies as a power-law function of the virial mass, 
whereas the logarithmic dispersion in the mass range of galaxy clusters is statistically consistent with being constant \citep[][]{Net07}. 
This means that the distribution of clusters on the mass-concentration diagram may be described by the following probability density 
function \citep[][]{Com07}
\begin{equation}\label{c_dis}
	f(c)\textrm{d}c=\frac{1}{\sqrt{2\pi}c \ln\Delta}\exp\Big[\frac{-(\ln c-\ln\mu)^2}{2(\ln\Delta)^2}\Big]\textrm{d}c,
\end{equation}
where $\mu=c(M_{0})(M_{\rm v}/M_{0})^{\alpha}$. In order to minimize the correlation between the slope and the normalization, the reference mass 
$M_{0}$ should be approximately the median value of the virial mass in the sample. In our case we assumed $M_{0}=5\times 10^{14}M_{\odot}$.

We carried out the likelihood analysis of the data shown in Fig.~\ref{c_mv_dia}. The likelihood function was the product of the probability density 
function (\ref{c_dis}) evaluated at the mass parameters of all clusters. The analysis was performed with the MCMC technique with three free 
parameters: the slope $\alpha$, the normalization $c(5\times 10^{14}M_{\odot})$ and the logarithmic dispersion $\ln\Delta$. We note that the latter 
parameter includes both the internal scatter of the theoretical mass-concentration relation and the inaccuracy of the measurement of the concentration 
parameter.

\begin{figure}
\begin{center}
    \leavevmode
    \epsfxsize=8cm
    \epsfbox[50 50 580 420]{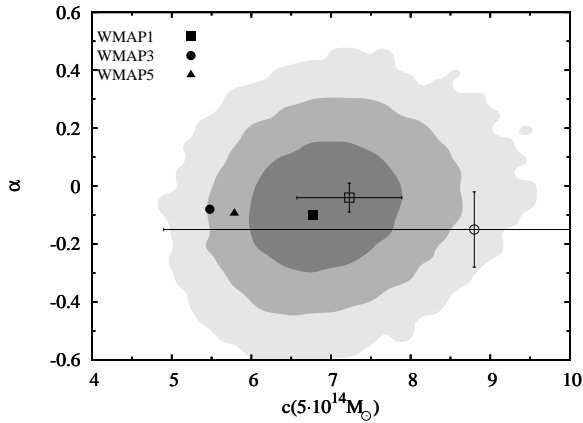}
\end{center}
\caption{Constraints on the parameters of the power-law fit to the mass-concentration data, the slope $\alpha$ and the 
normalization at the mass $5\times 10^{14}M_{\odot}$. The empty square and circle with error bars show the constraints on the parameters 
of the mass-concentration relation obtained by \citet[][]{Buo07} and \citet[][]{Com07}, respectively. Filled symbols indicate the 
calibrations of the mass-concentration relation from cosmological $N$-body simulations \citep[see][]{Macc08} with the parameters 
from three data releases of the WMAP satellite labelled by WMAP1, WMAP3 and WMAP5.}
\label{c_mv_MCMC}
\end{figure}

Fig.~\ref{c_mv_MCMC} shows the constraints on the slope and the normalization of the mass-concentration relation. Three contours correspond 
to the $1\sigma$, $2\sigma$ and $3\sigma$ credibility regions. Empty symbols with error bars indicate the constraints on the slope and the normalization 
of the $M_{\rm v}-c$ relation obtained by \citet{Buo07} from the analysis of X-ray clusters and by \citet{Com07} from the analysis of the mass parameters 
of galaxy clusters compiled from the literature. In both cases, the original results were properly adapted to the overdensity parameter assumed 
in our work as well as to the reference mass $M_{0}$. Our result agrees within errors with the constraints obtained by \citet{Buo07} and \citet{Com07}, 
although clearly a lower value of the normalization is preferred, $c(5\times 10^{14}M_{\odot})=6.9^{+0.6}_{-0.7}$. 

In order to compare our results with theoretical predictions we plotted in Fig.~\ref{c_mv_MCMC} the calibrations of the mass-concentration relation 
for relaxed dark matter haloes formed in cosmological $N$-body simulations \citep[see][]{Macc08}, where an appropriate change of the 
virial overdensity parameter was taken into account. We note that the reference simulations are purely dark matter simulations which 
do not take into account the steepening of dark matter profile induced by baryonic gas \citep[see][]{Gne04}. To account for this effect 
we increased all normalizations by $10$ per cent which is the value close to the upper limit of the bias in the concentration parameter 
found in hydrodynamical simulations of galaxy clusters \citep[see][]{Dol09,Duf10}. Each calibration 
(filled points in Fig.~\ref{c_mv_MCMC}) refers to one of three cosmological models given by the 
parameters from three data releases of the Wilkinson Microwave Anisotropy Probe (WMAP) satellite, WMAP1 \citep[][]{Sper03}, WMAP3 \citep[][]{Sper07}, 
and WMAP5 \citep[][]{Kom09}. The main difference between these three sets of cosmological parameters lies in the value of $\sigma_{8}$, the linear amplitude 
of density perturbations at $8$ $h^{-1}$Mpc scale at $z=0$, which ranges from $0.75$ for WMAP3, through $0.8$ for WMAP5, to $0.9$ for WMAP1. Keeping 
in mind that the normalization of the $M_{\rm v}-c$ relation is mostly sensitive to $\sigma_{8}$, the three values of $\sigma_{8}$ correspond to three 
normalizations labelled by WMAP1, WMAP3 and WMAP5 in Fig.~\ref{c_mv_MCMC}. From Fig.~\ref{c_mv_MCMC} one can see that our constraints on the 
normalization of the mass-concentration relation are fully consistent with $\sigma_{8}$ of WMAP1 cosmology and agrees only at $2\sigma$ and 
$\sim1\sigma$ level with that of WMAP3 and WMAP5, respectively. We confirm recent findings that the concentration parameters for galaxy 
clusters tend to be higher than those predicted by cosmological model with WMAP3 or WMAP5 parameters \citep[see][]{Buo07,Duf08}.

In order to place constraints on $\sigma_{8}$ using the limits for the normalization $c(M_{0})$ we adopted a semianalytical model for the 
mass-concentration relation proposed by \citet{Bul01}. According to this model the mean concentration parameter at redshift $z=0$ is inversely 
proportional to the scale factor $a_{\rm c}$ of halo collapse given by the following equation
\begin{equation}\label{bul_mod}
	\sigma(FM_{\rm v})=\frac{1.686}{D(a_{\rm c})},
\end{equation}
where $D(a)$ is the linear growth rate, $\sigma$ is the present linear rms density fluctuation at the mass scale $FM_{\rm v}$ and $F$ is a 
free parameter of this model. Assuming that the only factor determining the normalization $c(M_{0})$ at a fixed virial mass $M_{0}$ is the value 
of $\sigma_{8}$ we find that $\sigma_{8}\propto 1/D(a_{\rm c})$ and $c(M_{0})\propto 1/a_{\rm c}$. We used both proportionalities to convert 
$c(M_{0})$ into $\sigma_{8}$. In this conversion, we assumed the calibration from \citet{Macc08} for $\sigma_{8}=0.9$ and corrected 
for $10$ per cent bias upwards, i.e. $c(5\times 10^{14}M_{\odot})=6.8$ (see Fig.~\ref{c_mv_MCMC}), where $a_{\rm c}$ was calculated using 
equation (\ref{bul_mod}) with $F=0.001$ which is the recommended value for the mass regime of galaxy clusters \citep[][]{Bul01,Buo07}. 
As a result we expressed the constraints on the mean mass-concentration relation in terms of $\sigma_{8}$ parameter 
obtaining $\sigma_{8}=0.91_{-0.08}^{+0.07}$. We note that the difference between MAP value of $\sigma_{8}$ and its 
value from WMAP5 data release corresponds to the change of $c(M_{0})$ by $-14$ per cent. We cannot exclude that this difference 
may arise from systematic errors in the measurement of the concentration parameter,  observational selection biases 
or an insufficient value of the bias correcting the calibration of the mass-concentration relation for the presence of baryons.

Our constraint on the slope of the $M_{\rm v}-c$ relation is rather weak, i.e. $\alpha=-0.07\pm0.15$. It is consistent with a flat 
profile as well as with the slope obtained in cosmological simulations $(\alpha\sim -0.1)$. The best-fit value of the logarithmic dispersion 
parameter $\ln\Delta$ is $0.55$. The width of the light gray region in Fig.~\ref{c_mv_dia} indicates this value, whereas the black solid line 
and the dark gray area show the best-fit profile for the mean concentration $\mu(M_{\rm v})$ and the error of the normalization. Taking into account 
that the statistical scatter in the measurement of the concentration parameter is around $\sigma_{\ln\Delta}\approx 0.44$ \citep[see][]{Woj09}, we conclude 
that the logarithmic dispersion of an intrinsic scatter in the theoretical $M_{\rm v}-c$ relation is about $0.35$ which is in fair agreement with 
the results of cosmological simulations \citep[][]{Bul01,Net07,Macc08}. 
%As final remark we note that the results presented in this subsection 
%does not depend on a specific choice of the likelihood function \ref{c_dis}. 
%We checked that a simple linear fit in logarithmic scale with symmetrized 
%errors gives the same results.

\subsection{Mass scaling relations}

\begin{figure}
\begin{center}
    \leavevmode
    \epsfxsize=8cm
    \epsfbox[50 50 570 1170]{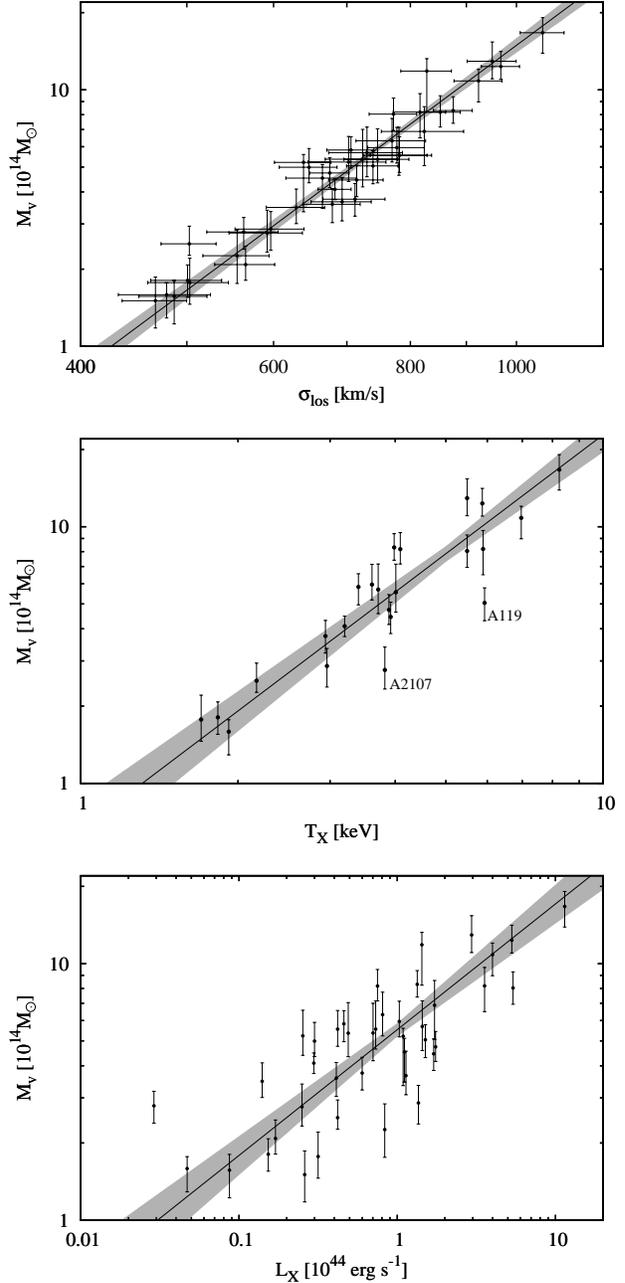}
\end{center}
\caption{Scaling relations between the virial mass and different mass proxies: the velocity dispersion ($\sigma_{\rm los}$), 
the X-ray luminosity ($L_{\rm X}$) and the X-ray temperature ($T_{\rm X}$). Solid lines show the best power-law fits and gray shaded 
areas represent the errors in both the slope and normalization.}
\label{mass_scaling}
\end{figure}

The virial mass of galaxy clusters scales with a number of different observables. Here we consider scaling relations with the velocity 
dispersion, the X-ray temperature and the X-ray luminosity (see Fig.~\ref{mass_scaling}). The virial mass, as listed in Table~\ref{tab_para}, 
comes from our analysis of $41$ velocity diagrams of nearby galaxy clusters. Velocity dispersions were calculated using all galaxies 
lying within the virial radii of the clusters. The errors were estimated by bootstrapping from the sample. We used the same 
kinematic data as in the proper analysis of velocity diagrams. The temperatures were obtained by \citet{Hor01} from the spectral analysis of 
data acquired by the ASCA satellite ($23$ clusters). The luminosities were taken from the catalogues of X-ray clusters observed by the 
ROSAT satellite in the energy band $0.1-2.4$ keV: the Northern ROSAT All-Sky (NORAS) galaxy cluster survey 
\citep[see][]{Boe00}, the ROSAT-ESO Flux Limited X-ray (REFLEX) galaxy cluster survey \citep[see][]{Boe04} and a complementary source of 
the data provided by \citet{Led03}. All luminosities were corrected for the assumed cosmological parameters.

\begin{table}
\begin{center}
\begin{tabular}{l|cc}
relation   & $\alpha$                     & $M_{0} [10^{14}M_{\odot}]$    \\
\hline
 & \\ 
$M_{\rm v}=M_{0} \Big[\sigma_{\rm los}/(750 \textrm{ km s$^{-1}$})\Big]^{\alpha}$ & $3.17\pm0.13$ & $5.99\pm0.17$ \\
$M_{\rm v}=M_{0} \Big[L_{\rm X}/(10^{44} \textrm{ erg s}^{-1})\Big]^{\alpha}$ & $0.49\pm0.05$ & $5.53\pm0.33$ \\
$M_{\rm v}=M_{0} \Big[T_{\rm X}/(5\textrm{ keV})\Big]^{\alpha}$ & $1.54\pm0.12$ & $7.85\pm0.55$ \\
\end{tabular}
\caption{Best-fit parameters of the power-law approximation for the three different mass scaling relations.}
\label{scale_par}
\end{center}
\end{table}

All scaling relations seen in Fig.~\ref{mass_scaling} are well approximated by a power-law function. Table~\ref{scale_par} lists the best-fit 
slope and normalization for each of them. Parameters of the scaling relations were obtained by fitting a line to the data in logarithmic 
scales. We used symmetrized logarithmic errors given by $(x_{2}-x_{1})/(2x_{0})$, where $x_{1}$ and $x_{2}$ are the boundaries of the 
$1\sigma$ credibility range of $x$ variable, and $x_{0}$ is the most probable value. The best-fit slope and normalization were calculated using 
the bisector method \citep[][]{Iso90}. The pivot points of $\sigma_{\rm los}$, $L_{\rm X}$ and $T_{\rm X}$ given in Table~\ref{scale_par} were 
taken to be approximately the median values of the corresponding variables. This minimized the correlation between the slope and normalization. 
The errors were evaluated by bootstrap resampling.

The best-fit scaling relations were plotted in Fig.~\ref{mass_scaling}. Solid lines are the best power-law fits and 
the shaded areas indicate the errors in both the slope and normalization. Both the temperature and the velocity dispersion 
appear to correlate very well with the virial mass. We note, however, that the a tight correlation with the velocity 
dispersion occurs mainly due to the fact that the virial masses were derived from the same kinematic data that were used 
to evaluate velocity dispersions. Two clusters, A119 and A2107, lie below the mean 
mass-temperature relation. This discrepancy is probably caused by the fact that these clusters are less relaxed compared to 
the other ones in our sample: A119 exhibits distinct substructering \citep{Sch01}, whereas A2107 has a high-velocity 
($\sim 200$ km s$^{-1}$) cD galaxy that may suggest some merging activity. The correlation between X-ray luminosities and the virial 
masses is weaker and confirms a well-known fact that the X-ray luminosity is not a robust mass proxy. Nevertheless, we 
note that our constraint on the slope of the $M_{\rm v}-L_{X}$ relation is in fair agreement with other measurements 
available in the literature ranging between $0.5$ and $0.7$ \citep[see e.g.][]{Pop05,Lop09}.

The slopes of the mass scaling relations with the X-ray temperature as well as with the velocity dispersion are fully 
consistent with the predictions of the virial theorem which states that $M_{\rm v}\propto \sigma^3\propto T_{\rm X}^{3/2}$ 
\citep[][]{Bry98}. The slope of the mass-temperature relation is in excellent agreement with the most recent 
estimates which yield values between $1.5$ and $1.6$ \citep[see e.g.][]{Vik06}. The normalization of the same relation, 
when converted to the overdensity $\Delta_{c}=500$, is $(4.7\pm 0.3)\times 10^{14}M_{\odot}$ and agrees with that 
derived from high-resolution hydrodynamical simulations \citep[see e.g.][]{Bor04}. It is consistent with the 
normalization of X-ray derived mass-temperature relation \citep[see e.g.][]{Arn05,Vik06} and is lower by $\sim20$ 
per cent than mass-temperature normalization determined from weak lensing masses \citep[see e.g.][]{Ped07}.

Our analysis provides an independent test of the scaling relations with X-ray observables on the scale of the virial overdensity 
that is usually not used for X-ray clusters. Since X-ray flux is detected in the very central part of clusters, a typical scale of 
the mass probed in X-ray observations is between $M_{200}$ and $M_{500}$. This mass scale is also commonly used to study X-ray 
scaling relations. Our analysis shows that the mass scaling relations both with $T_{\rm X}$ and $L_{\rm X}$ may be extended to 
the virial mass with the mean overdensity $\sim100\rho_{c}$.

\subsection{The anisotropy of galaxy orbits: individual clusters}

\begin{figure}
\begin{center}
    \leavevmode
    \epsfxsize=8cm
    \epsfbox[50 50 580 430]{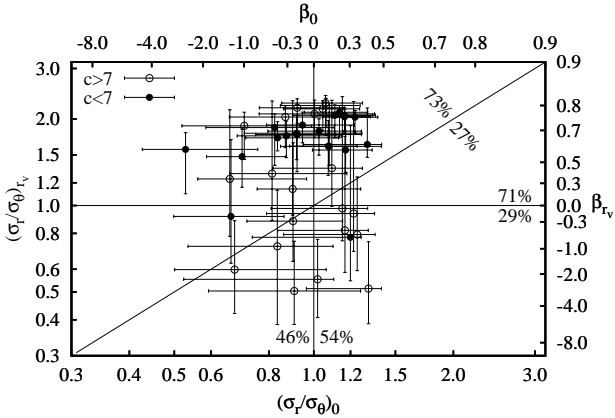}
\end{center}
\caption{The orbital anisotropy of galaxies in $41$ galaxy clusters. Empty and filled symbols refer to the clusters with the concentration parameter 
above or below the median concentration in the sample ($c=7$). Numbers indicate fractions of clusters in domains of the diagram 
defined by the lines of isotropic velocity distribution ($\beta_{0}=0$ or $\beta_{r_{\rm v}}=0$) and the family of flat anisotropy 
profiles ($\beta_{0}=\beta_{r_{\rm v}}$).}
\label{beta_sca}
\end{figure}

Fig.~\ref{beta_sca} shows the constraints on the orbital anisotropy of galaxies in the cluster centre ($\beta_{0}$) and at the virial sphere 
($\beta_{r_{\rm v}}$). Although \citet{Woj09} showed that the estimate of the anisotropy in the outer part of the cluster appears to be 
much less accurate than in the central part, we think that some interesting conclusions may be drawn from studying the overall 
distribution of the results obtained for individual clusters. The estimates of the central anisotropy for all clusters are statistically 
consistent with $\beta_{0}=0$. Galaxy orbits at the virial sphere of the clusters are noticeably radially biased with a typical 
anisotropy of around $(\sigma_{r}/\sigma_{\theta})_{r_{\rm v}}=1.7$. We also note that there is a clear tendency of the anisotropy profile 
to increase with radius: over $70$ per cent of the results lie above the line of flat anisotropy profiles indicated by the diagonal. 
On the other hand, some clusters exhibit flat $\beta$ profile with an isotropic velocity dispersion tensor. 

Clusters with different degrees of dark matter concentration tend to populate different regions of the parameter space shown in 
Fig.~\ref{beta_sca}. In particular, low concentration clusters ($c<7$) exhibit more radially biased galaxy orbits at the virial 
sphere than clusters with higher value of the concentration ($c>7$). Furthermore, the former clusters seem to constitute a very 
homogeneous class in terms of the anisotropy profile. The orbital anisotropy of galaxies in these clusters seems to grow with radius between 
two universal asymptotes: $\sigma_{r}/\sigma_{\theta}\approx 1$ in the cluster centre and $\sigma_{r}/\sigma_{\theta}\approx 1.8$ at the virial 
radius. In the case of the high concentration clusters, the anisotropy profiles do not show any signatures of such universality. The 
anisotropy at the virial radius in these clusters ranges from $(\sigma_{r}/\sigma_{\theta})_{r_{\rm v}}\sim 1$ to 
$(\sigma_{r}/\sigma_{\theta})_{r_{\rm v}}\sim 2$. Nevertheless, it is interesting to note that $85$ per cent of the clusters for which 
constraints on the anisotropy parameters are statistically consistent with the fully isotropic model ($\beta_{0}=\beta_{r_{\rm v}}=0$) 
are high concentration clusters.

Keeping in mind that the degree of mass concentration in galaxy clusters is related to the state of dynamical evolution, 
we suggest that the dichotomy between the low and high concentration clusters observed in Fig.~\ref{beta_sca} may be likely 
the result of the cluster evolution. In this scenario galaxy orbits of less evolved clusters (low concentration parameter) would 
be dominated by radial motions in their outer parts (filled symbols in Fig.~\ref{beta_sca}), whereas for more evolved clusters 
galaxy orbits tend to be more isotropic both in the cluster centre as well as in the vicinity of the virial sphere. 

At the early stage of cluster evolution one can expect an excess of radial orbits around the virial sphere. 
This excess is mostly the effect of radial infall of galaxies onto the clusters. Due to the relaxation process the initial orbits 
of infalling galaxies are expected to become more isotropic with time so that the most evolved clusters are characterized by more isotropic 
distribution of galaxy orbits. A similar mechanism of the isotropization of galaxy orbits in clusters was pointed out by \cite{Biv04}. 
Studying the orbital structure in ENACS (ESO Nearby Abell Cluster Survey) clusters, \citet{Biv04} showed that early type galaxies 
(ellipticals) are characterized by more isotropic orbits than spirals for which $\sigma_{r}/\sigma_{\theta}$ at the virial sphere is 
around $1.7$. Since the former galaxy population is more evolved and relaxed than the latter, this orbital segregation implies that 
a mechanism of the orbital isotropization must operate during cluster evolution.

\subsection{The anisotropy of galaxy orbits: a joint analysis}

In order to estimate the typical anisotropy profile of galaxy orbits in the clusters of our sample we carried out a joint 
MCMC analysis of all velocity diagrams. Following \cite{Woj09} we defined the likelihood function as
\begin{eqnarray}
	f(M_{s,1},\dots,M_{s,n},r_{s,1},\dots,r_{s,n},\beta_{0},\beta_{\infty})=\nonumber\\
	\prod_{j=1}^{n}\prod_{i=1}^{N_{\rm mem}(j)}f_{\rm los}(R_{j,i},v_{{\rm los},j,i}|\{M_{s,j},r_{s,j},
	\beta_{0},\beta_{\infty}\})		\label{like_ndiagram},
\end{eqnarray}
where $j$ and $i$ are respectively the reference numbers of a clusters and a galaxy of the $j$-th velocity diagram. The two 
parameters of the anisotropy profile, $\beta_{0}$ and $\beta_{\infty}$, are common to all clusters and define the anisotropy 
profile in cluster population, whereas parameters of the mass profile fix phase-space units for individual clusters. 
We used the constraints on the parameters of the mass profile, $M_{\rm s}$ and $r_{\rm s}$, in the 
form of Gaussian prior, where the means and the dispersions were given by MAP values and symmetrized errors corresponding 
to the $1\sigma$ credibility ranges.

\begin{figure}
\begin{center}
    \leavevmode
    \epsfxsize=8cm
    \epsfbox[50 50 580 430]{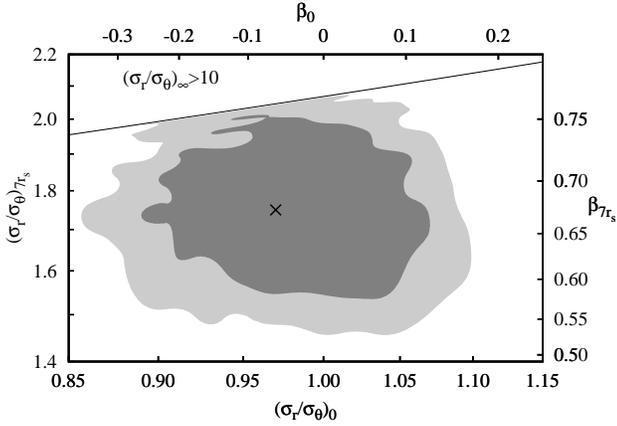}
\end{center}
\caption{Constraints on the parameters of the anisotropy profile of galaxy orbits from the joint analysis of $41$ nearby galaxy clusters. 
The gray shading shows the $1\sigma$ and $2\sigma$ confidence regions of the marginal probability distribution and the cross indicates the MAP 
value. The black solid line marks the boundary of the parameter space excluded from the analysis by the prior 
$(\sigma_{r}/\sigma_{\theta})_{\infty}<10$.}
\label{beta_41-joint}
\end{figure}

Fig.~\ref{beta_41-joint} shows the result of a joint analysis in the form of the $1\sigma$ and $2\sigma$ credibility contours of 
the marginal probability distribution. In order to avoid extrapolation beyond the virial sphere we converted $\beta_{\infty}$ 
parameter into the anisotropy at $7r_{\rm s}$ which is a typical scale of the virial radius in the cluster sample. The black solid line 
indicates the boundary of the parameter space excluded from the analysis by the prior $(\sigma_{r}/\sigma_{\theta})_{\infty}<10$. 
The joint analysis confirms that galaxy orbits are typically isotropic in the cluster centres, 
$(\sigma_{r}/\sigma_{\theta})_{0}=0.97\pm0.04$, and radially anisotropic at the virial sphere, 
$(\sigma_{r}/\sigma_{\theta})_{7r_{\rm s}}=1.75^{+0.23}_{-0.19}$. The degree of the anisotropy at the virial sphere agrees with the 
value of the orbital anisotropy determined for late type galaxies which dominate in the outer parts of galaxy clusters \citep{Biv04}. 
We do not confirm the result obtained by \citet{Mar00} who concluded that velocity distribution in galaxy clusters is consistent 
with isotropic orbits. The reason of this discrepancy may lie in that fact that \citet{Mar00} assumed a flat profile of the anisotropy. 
Taking into account that the central anisotropy is much better constrained than the anisotropy in the outer part of 
a cluster \citep[][]{Woj09}, one may expect that the estimate of the global anisotropy is dominated by its central value. Consequently, 
the final estimate of the anisotropy is closer to its central rather than the outermost value. 
%Interestingly, the same situation occurs in 
%the analysis of velocity moments of individuals clusters. We checked that assuming flat anisotropy profile our analysis yields 
%$\beta\approx0.17$ or $\sigma_{\rm r}/\sigma_{\theta}\approx1.1$.
%Assuming constant anisotropy, \citet{Lok06} and \citet{Woj07} showed that galaxy 
%kinematics of $\sim 20$ galaxy clusters is consistent with isotropic orbits.

\begin{figure}
\begin{center}
    \leavevmode
    \epsfxsize=8cm
    \epsfbox[50 50 580 430]{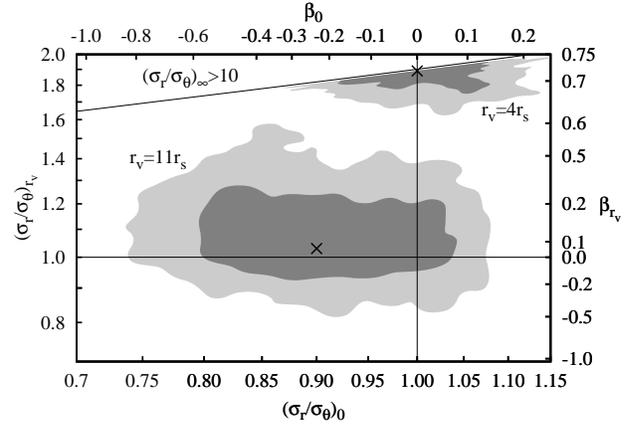}
\end{center}
\caption{Constraints on the parameters of the anisotropy profile of galaxy orbits from the joint analysis of low concentration 
clusters (with $c<7$ and the median virial radius $r_{\rm v}\approx4r_{\rm s}$) and high concentration clusters (with $c>7$ and the median 
virial radius $r_{\rm v}\approx11r_{\rm s}$. The gray shading shows the $1\sigma$ and $2\sigma$ confidence regions of the marginal probability 
distribution and the crosses indicate the MAP values. The black solid line marks the boundary of the parameter space excluded from the 
analysis by the prior $(\sigma_{r}/\sigma_{\theta})_{\infty}<10$.}
\label{beta_41-joint_bim}
\end{figure}

The size of the credibility regions in Fig.~\ref{beta_41-joint} does not reflect the scatter of the anisotropy profiles in the 
cluster sample. It rather describes the accuracy of parameter estimation under the assumption that the cluster sample is 
homogeneous in terms of the anisotropy profile. In order to illustrate the degree of the internal scatter of the anisotropy parameters 
within our cluster sample we repeated the joint analysis for two separate cluster subsamples, clusters with the concentration parameters 
above ($c>7$) and below ($c<7$) the median (see Fig.~\ref{beta_41-joint_bim}). The choice of these subsamples is motivated by the observed segregation 
of low and high concentration clusters on the plane of the anisotropy parameters (see Fig.~\ref{beta_sca}). Our result confirms the conclusion 
from the previous subsection that low concentration clusters have galaxies on clearly radially biased orbits at the virial sphere, 
whereas galaxy orbits for high concentration clusters tend to be isotropic at all radii within the virial sphere.

\section{Comparison with other methods}

For a number of galaxy clusters parameters of the mass profile have already been determined with other methods. In this section, 
we use these estimates to compare with the results obtained with our method. This comparison allows us to check the credibility of 
our approach with respect to other methods of mass inference for galaxy clusters.

\begin{figure}
\begin{center}
    \leavevmode
    \epsfxsize=8cm
    \epsfbox[50 50 580 790]{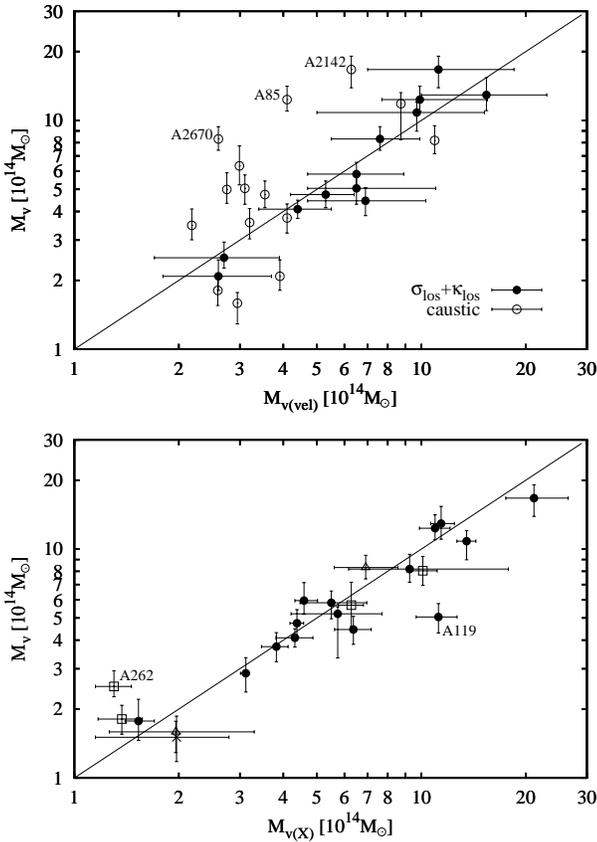}
\end{center}
\caption{Comparison between the virial mass obtained with the method of the projected phase-space analysis ($M_{\rm v}$) and 
other methods based on the X-ray observations ($M_{\rm v(X)}$, bottom panel), the Jeans analysis of velocity moments and 
the caustic technique ($M_{\rm v(vel)}$, top panel). The sources of data for subsequent point types are the following: 
filled circles in the top panel: \citet{Lok06,Woj07}; empty circles in the top panel: \citet{Rin03,Rin06}; filled circles 
in the bottom panel: \citet{Rei02}; empty squares: \citet{Vik06}; empty triangle: \citet{San03}; cross: \citet{Poi05}. The 
black solid line represents equality between the mass estimates.}
\label{comp_Mv}
\end{figure}

Fig.~\ref{comp_Mv} shows the comparison between the virial mass estimated in our analysis and with other methods available for 
nearby galaxy clusters. The bottom panel includes the mass estimates obtained from the X-ray observations. 
It comprises both the results of a detailed analysis of the temperature profiles \citep{Poi05,Vik06} and 
the mass estimates obtained under the assumption that the intracluster gas is isothermal \citep{Rei02} or is linked to the temperature 
via a polytropic equation of state \citep{San03}. The top panel presents the mass estimates from galaxy kinematics including 
the Jeans analysis with the velocity dispersion $\sigma_{\rm los}$ and 
the kurtosis $\kappa_{\rm los}$ \citep{Lok06,Woj07}, and the caustic method \citep{Rin03,Rin06}. 
We note that all mass estimates were adopted to the virial overdensity assumed in this work. In the conversion between 
masses corresponding to different overdensities, we used the same mass profile as assumed by the authors of a given data analysis. 
Apart from the analysis carried out by \citet{Rei02} and \citet{San03} who used the mass profile derived from the equation of hydrostatic 
equilibrium with a fixed parametrization of the gas density and the temperature profile, the mass profile was parametrized by the 
NFW formula (\ref{NFW}).

Our mass determinations agree within errors with the estimates based on X-ray analysis. There are two clusters, 
A119 and A262, for which the difference between mass estimates exceeds significantly the $1\sigma$ level. The reason of this discrepancy 
seems to be clear for A119, since this is one of the least relaxed clusters of our sample \citep{Sch01}. On the other hand, 
an analogous discrepancy for A262 remains unclear, since there is no observational evidence which would indicate that this 
cluster is not in equilibrium. We checked that both the X-ray mass and the mass inferred from galaxy kinematics, as displayed 
in Fig.~\ref{comp_Mv}, are consistent with the corresponding result obtained by independent authors \citep[][]{Lok06,Gas07}. 
This suggests that this difference in probably a consequence of a generic tension between kinematic and X-ray data.

The virial masses obtained in our work agree very well with those determined in the analysis of velocity moments. This fact is fully 
understandable, since our approach to data modelling is an extension of the formalism based on velocity moments. We note that the 
$1\sigma$ credibility ranges obtained with our method are noticeably narrower than those resulting from the Jeans analysis and are 
comparable to the errors of the X-ray masses.

The caustic method provides mass estimates which marginally agree with the virial masses determined with our approach. The caustic mass 
correlates with the virial mass from our analysis, but is considerably scattered around the line of equal mass estimates and tends to 
be underestimated with respect to our mass determination. For three cases, A85, A2142 and A2670, this effect is most prominent: 
the caustic virial mass is $\sim 3$ smaller than its counterpart from our analysis. We checked that our measurements of the virial 
masses for these clusters agree with the estimates from the X-ray observations yielding $\sim 1.1\times10^{15}M_{\odot}$ for A85 
\citep{Rei02,Dur05}, $\sim 2.1\times10^{15}M_{\odot}$ for A2142 \citep{Rei02} and $\sim 7\times10^{14}M_{\odot}$ for A2670 
\citep{San03}. This confirms that the caustic masses of these three clusters are likely underestimated.

\begin{figure}
\begin{center}
    \leavevmode
    \epsfxsize=8cm
    \epsfbox[50 50 580 410]{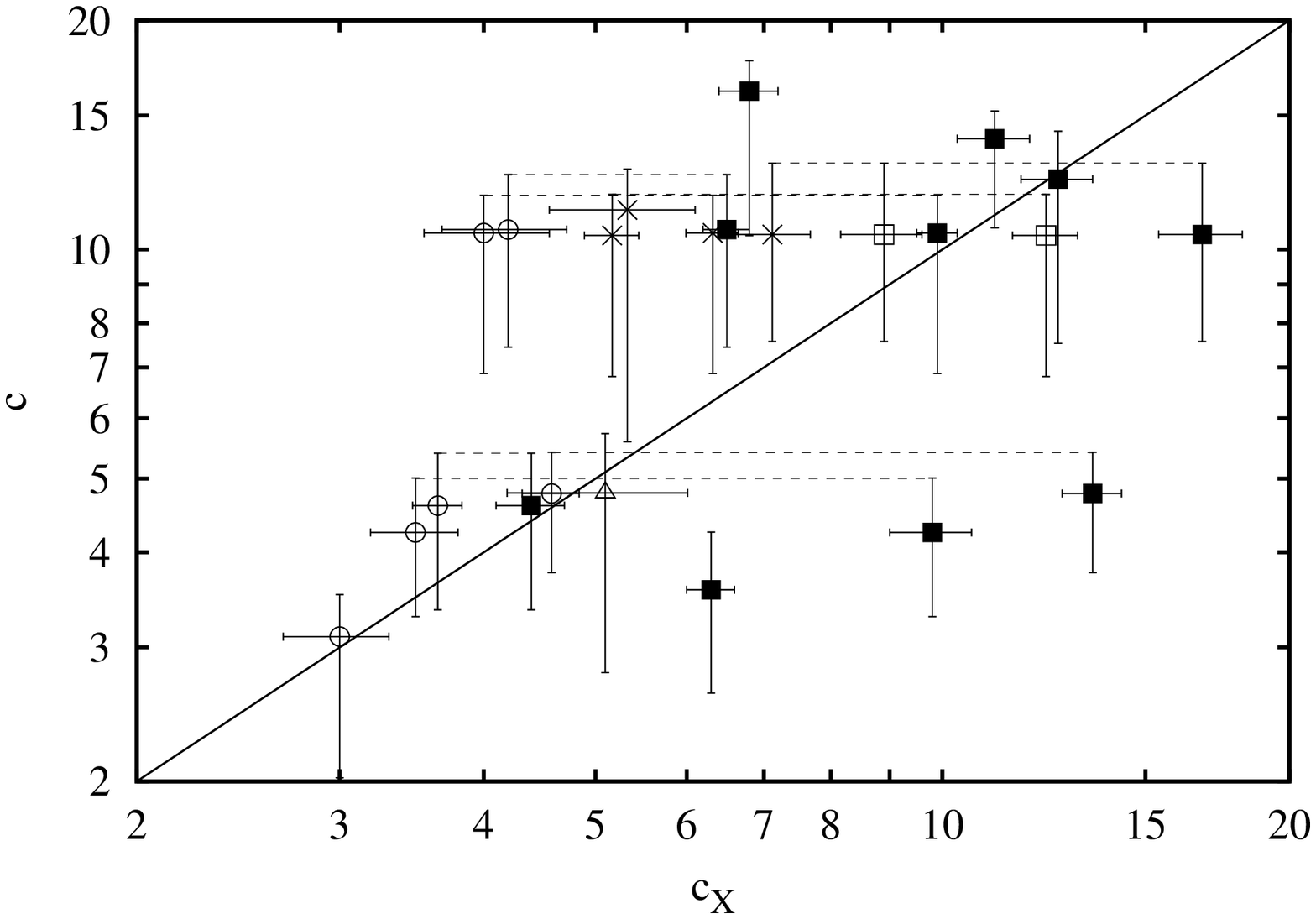}
\end{center}
\caption{Comparison between the concentration parameter obtained with the method of the projected phase-space analysis ($c$) 
and from the X-ray observations ($c_{\rm X}$). The sources of data for subsequent point types are the following: 
filled squares: \citet{Xu01}; empty squares: \citet{Gas07}; crosses: \citet{Vik06}; circles: \citet{Ett02}; triangle: 
\citet{Poi05}. The black solid line represents equality between both estimates of the concentration parameter. The dashed lines 
connect data for the same clusters.}
\label{comp_con}
\end{figure}

Fig.~\ref{comp_con} shows the comparison between the concentration parameters obtained with our method and from the X-ray observations 
of the corresponding clusters \citep[see][]{Xu01,Ett02,Vik06,Gas07}. 
%For this comparison we chose only clusters for which constraints 
%on the concentration parameters were comparable or better than our results. 
We note that some clusters (A85, A119, A262, A496, A1795, A3571 and MKW4) have multiple representations on this diagram corresponding 
to the estimates obtained by different authors. Since for some clusters these estimates do not agree within the errors, we decided to 
plot all results. The points referring to the same cluster were connected with a horizontal dashed line. As for the comparison of the 
virial masses, the concentration parameters were properly converted to the virial overdensity assumed in this work.

The concentration parameter determined with our method correlates with that inferred from the X-ray data, although the scatter is 
significantly larger than for the virial mass. Assuming a power-law relation between both estimates, 
$c=c_{7}(c_{\rm X}/7)^{\alpha}$, we obtained the following constraints on the slope and the normalization: $\alpha=1.04\pm0.4$ and 
$c_{7}=8.2\pm1.0$, where the fit was performed in the same manner as described in subsection 3.3. Although the data sample is not 
statistically uniform so we cannot draw any firm conclusion, we point out that X-ray derived concentrations tend to be smaller by 
$\sim15$ per cent than the concentrations inferred from galaxy kinematics.

%that the value of the normalization implies that our method typically 
%overestimates the concentration parameter by around $15$ per cent with respect to the X-ray data. Assuming that the inference of the 
%concentration parameter from the X-ray data is more reliable than by our method, it means that the normalization of the mass-concentration 
%relations obtained in subsection 4.2 may be biased upwards by $\sim15$ per cent. We note that this bias is sufficient to explain the difference 
%between $\sigma_{8}$ inferred from the mass-concentration relation and the value provided by WMAP5.

%This, however, stands in contradiction to the analysis carried out \citet{Woj09} who concluded 
%from the series of tests on mock kinematic data that the concentration parameter determined in the projected phase-space analysis is 
%unbiased. Solving this problem demands probably more tests of consistency between both methods from the point of view of the estimation 
%of the concentration parameter.

\section{Self-consistency tests}

\subsection{Completeness}

An important factor that may influence the quality of our analysis is the completeness of spectroscopic data. In order 
to check this effect we carried out the following test. If the data are incomplete one may expect 
that the parameters estimated from velocity diagrams truncated at different projected radii will be 
different. To see if this is the case we repeated the analysis for all clusters with three different values of the maximum 
projected radius: $R_{\rm max}=1$ Mpc, $R_{\max}=1.7$ Mpc and $R_{\max}=r_{\rm v}$, where $r_{\rm v}$ is the virial radius 
taken from the first analysis with $R_{\rm max}=2.5$ Mpc. Table~\ref{tab_compl} shows the comparison between 
parameter estimates obtained for different sizes of the aperture. Each row contains the logarithmic mean and 
dispersion of the relative residuals for the MAP values, where the reference values of the parameters come from 
the analysis with $R_{\rm max}=2.5$ Mpc (see Table~\ref{tab_para}).

\begin{table}
\begin{center}
\begin{tabular}{l|ccc}
 & $1$ Mpc & $1.7$ Mpc & $r_{\rm v}$ \\
\hline
$M_{\rm s}$ & $0.001\pm0.140$ & $0.015\pm0.056$ & $0.017\pm0.049$ \\
$r_{\rm s}$ & $-0.030\pm0.171$ & $0.012\pm0.094$ & $0.020\pm0.077$ \\
$M_{\rm v}$ & $0.038\pm0.173$ & $-0.010\pm0.101$ & $-0.019\pm0.082$ \\
$c$         & $0.025\pm0.105$ & $0.008\pm0.042$ & $0.003\pm0.033$ \\
$(\sigma_{r}/\sigma_{\theta})_{0}$ & $-0.035\pm0.101$ & $-0.008\pm0.038$ & $-0.003\pm0.026$ \\
$(\sigma_{r}/\sigma_{\theta})_{\infty}$ & $-0.112\pm0.588$ & $0.001\pm0.253$ & $-0.007\pm0.219$ \\
\end{tabular}
\caption{Statistics of the relative residuals of the model parameters from the analysis of $41$ velocity diagrams with three different 
truncation radii, $R_{\rm max}$. Each row lists the means and dispersions in the cluster sample for the relative residuals 
defined by $\log(X_{R_{\rm max}}/X_{2.5{\rm Mpc}})$, where $X_{R_{\rm max}}$ is the MAP value of the parameter obtained in the analysis 
of velocity diagrams truncated at $R_{\rm max}$.}
\label{tab_compl}
\end{center}
\end{table}

We see that there is no trend in the mean value of the residuals. One may notice a negative bias in the outer asymptotic 
value of the anisotropy profile for $R_{\rm max}=1$ Mpc. This is, however, a consequence of the fact that using a 
smaller aperture we probe the inner part of the clusters for which the orbits are more isotropic. We also note that the 
logarithmic dispersions which describe the scatter of the relative residuals are smaller or comparable to the relative 
errors associated with the $1\sigma$ credibility ranges listed in Table~\ref{tab_para}. This property together with the 
stability of the mean values means that our results are not sensitive to the change of the truncation radius of velocity 
diagrams. This excludes the possibility that the data may be subject to a significant incompleteness effect.

\subsection{Mass-to-number density}

\begin{figure}
\begin{center}
    \leavevmode
    \epsfxsize=8cm
    \epsfbox[50 50 580 430]{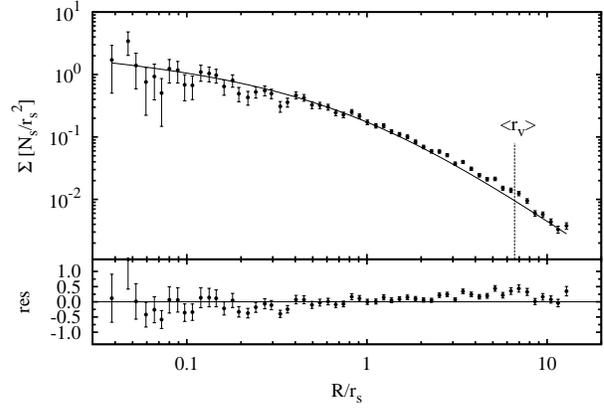}
\end{center}
\caption{Surface number density of galaxies in $41$ galaxy clusters combined into one. The black solid line is the projected 
NFW profile. The vertical line indicates a typical scale of the virial radius.}
\label{sigma_test}
\end{figure}

As a second test of self-consistency between the assumptions of our analysis and the data we show in Fig.~\ref{sigma_test} 
the surface number density of galaxies in all clusters combined into one. The surface density was evaluated 
in radial bins equally spaced in logarithmic scale, where the universal scaling of radii by $r_{\rm s}$ was adopted ($N_{\rm s}$ 
is the number of galaxies within the $R<r_{\rm s}$ aperture). The points above $3r_{\rm s}$ were corrected 
for the incompleteness of velocity diagrams which occurs due to the fact that the value of $R_{\rm max}/r_{\rm s}$ 
is not universal in the cluster sample. As a line of reference we plotted the projected NFW profile (black solid line). 
Apart from radii above $\sim 3r_{\rm s}$, where some deviations between the model and the data occur mostly due to the problem 
of incompleteness, galaxy density is well approximated by the NFW profile. Our data for small radii do not reveal a signature 
of a core in galaxy distribution, as advocated by \citet{Pop07}. Although our analysis concerns the projected densities, we may 
conclude from Fig.~\ref{sigma_test} that the assumption of our phase-space density model about a constant mass-to-number density 
is also satisfied. This confirms the observational fact that has already been discussed in the literature 
\cite[see e.g.][]{Car97,Biv03}.

\section{Summary and conclusions}

In this paper we presented the analysis of $41$ nearby ($z<0.1$) relaxed galaxy clusters. Using 
the method based on the modelling of velocity diagrams in terms of the projected phase-space density, as 
described in section 3, we constrained parameters of the total mass and anisotropy profiles in these clusters. 
Parameter estimation was carried out in the framework of Bayesian data analysis with the use of the MCMC 
technique.

We found that the concentration parameter is weakly correlated with the morphological type of the clusters 
as well as with the presence of a cool core. We interpret this fact as a signature of cluster evolution from 
the state with a flatter density profile (low concentration parameter) to one with a steeper density profile (higher 
concentration parameter). Our explanation is consistent with the recent results of cosmological simulations 
which predict steeper density profiles for more dynamically evolved clusters \citep[][]{Macc08,Dol09,Duf10}.

The virial mass of the clusters correlates very well with the velocity dispersion, the X-ray temperature and the X-ray 
luminosity of the clusters. The slopes of the mass scaling relations agree with the estimates found in the 
literature and in the case of the velocity dispersion and X-ray temperature are consistent with the prediction from the virial 
theorem. We used constraints on the virial mass and the concentration parameter to test the most up-to-date estimates of the 
$\sigma_{8}$ parameter 
%(linear amplitude of density perturbations at $8$ $h^{-1}$Mpc scale at $z=0$) 
against the observational relation between the virial mass and the concentration parameter. Comparing the normalization of 
the mass-concentration relation for our cluster sample with the calibration from cosmological simulations we found that our 
data favours a rather high value, $\sigma_{8}=0.9_{-0.08}^{+0.07}$. Like \citet{Buo07}, we found an excellent agreement with 
$\sigma_{8}$ from the 1st data release of the WMAP satellite.

Our model of the projected phase-space density allowed us to constrain the profile of the orbital anisotropy in 
clusters. We showed at high significance level that galaxy orbits are isotropic in the cluster centres and radially 
anisotropic at the virial sphere, with $\beta\approx0.65$ ($\sigma_{r}/\sigma_{\theta}\approx 1.7$). This finding is 
consistent with the results obtained by \citet{Biv04} for late type galaxies in nearby galaxy clusters and generalizes 
a number of estimates of the global anisotropy of galaxy orbits in clusters \citep[see e.g.][]{Lok06,Woj07,Mar00}. 

Our constraints on the asymptotic values of the anisotropy parameter implies a growth of the anisotropy from 
$\beta\approx0$ at the cluster centre to $\beta\approx 0.4$ at the scale radius. Interestingly, this result 
agrees fairly well with recent estimates of the anisotropy for dark matter in galaxy clusters \citep{Hos09,Han07}.

The orbital anisotropy at the virial sphere of the low concentration clusters tends to be more radially biased than for 
the clusters of high concentration parameter. We suggest that this tendency is a consequence of a change in the galaxy 
orbits from radially biased at the early stage of cluster evolution to more isotropic for more dynamically evolved 
clusters. This scenario is supported by the studies of the orbital anisotropy for different galaxy populations which 
reveals a similar dichotomy between early and late type galaxies representing more and less relaxed galaxy populations 
in the clusters \citep[see][]{Biv04}. 

\section*{Acknowledgments}

RW wishes to thank G. Mamon, S. Gottl\"ober, A. Schwope and G. Lamer for discussions. RW is grateful for the hospitality of 
Institut d'Astrophysique de Paris and Astrophysikalisches Institut Potsdam where parts of this work were done. This research 
has made use of the NASA/IPAC Extragalactic Database (NED) which is operated by the Jet Propulsion Laboratory, California 
Institute of Technology, under contract with the National Aeronautics and Space Administration. This work was partially 
supported by the Polish Ministry of Science and Higher Education under grant NN203025333 as well as by 
the Polish-French collaboration program of LEA Astro-PF. RW acknowledges support from the START Fellowship for Young 
Researchers granted by the Foundation for Polish Science.

\end{document}